\documentclass[a4paper,12pt,twoside,onecolumn,number,review]{ArXiv}

\usepackage{times,url,graphicx,epsfig}
\usepackage{amsmath,amsthm,amssymb}
\usepackage[ruled,vlined,linesnumbered,algoruled]{algorithm2e}
\usepackage{setspace}
\usepackage{booktabs}
\usepackage{fancyhdr}

\usepackage{hhline}
\usepackage{booktabs}
\usepackage{float}
\usepackage[caption=false]{subfig}
\usepackage{threeparttable}
\usepackage{lscape}
\usepackage[usenames, dvipsnames]{color}

\newtheorem {theorem} {Theorem}
\newtheorem {definition} {Definition}
\newtheorem {lemma} {Lemma}

\newcommand {\dc}[1]     {{\mathcal C}^{\mathbb Z}(#1)}       
\newcommand {\dco}[2]    {{\mathcal C}_{#1}^{\mathbb Z}(#2)}  
\newcommand {\dd}[1]     {{\mathcal D}^{\mathbb Z}(#1)}       
\newcommand {\du}[1]     {{\mathcal D}_\cup^{\mathbb Z}(#1)}  
\newcommand {\dct}[2]    {{\mathcal C}_{#1}^{\mathbb Z}(#2)}  

\newcommand {\zz}        {{\mathbb Z}^2}

\newcommand {\zzz}       {{\mathbb Z}^3}
\newcommand {\sm}        {\smallsetminus}

\newcommand {\inzp}      {\in {\mathbb Z}^+}

\renewcommand {\sp}[1]   {{\mathcal S}^{\mathbb Z}(#1)}      
\newcommand {\he}[1]     {{\mathcal H}^{\mathbb Z}(#1)}      
\newcommand {\sr}[1]     {{\mathcal S}_\cup^{\mathbb Z}(#1)} 
\newcommand {\hr}[1]     {{\mathcal H}_\cup^{\mathbb Z}(#1)} 

\renewcommand {\ss}[1]   {{\mathbf S}^{\mathbb Z}(#1)}       
\newcommand {\sa}[1]     {{\mathbf S}_\cup^{\mathbb Z}(#1)}  
\newcommand {\hrs}[1]    {{\mathbf H}_\cup^{\mathbb Z}(#1)}  

\newcommand {\ad}[1]     {{{\mathcal A}^{{\mathbb Z}^2}}(#1)} 
\newcommand {\ap}[2]     {{\mathcal A}_{#1}^{{\mathbb Z}^2}(#2)}
\newcommand {\as}[1]     {{\mathcal A}^{{\mathbb Z}^3}(#1)}   
\newcommand {\aS}[1]     {{\mathbf A}^{{\mathbb Z}^3}(#1)}    
\newcommand {\al}[1]     {{\mathbf L}^{{\mathbb Z}^3}_{(#1)}} 
\newcommand {\ac}[1]     {{\mathbf C}^{{\mathbb Z}^3}_{(#1)}} 

\newcommand {\hf}[1]     {{\mathcal H'}^{\mathbb Z}(#1)}      

\newcommand {\ipc}[1]    {{\underline{P}}_{h,#1}} 
\newcommand {\spc}[1]    {{\overline{P}}_{h,#1}} 
\newcommand {\ips}[1]    {{\underline{\mathbf P}}_{h,#1}} 
\newcommand {\sps}[1]    {{\overline{\mathbf P}}_{h,#1}} 
\newcommand {\pv}[2]     {{\mathbf P}_{#1,#2}} 
\newcommand {\fa}[1]     {{\mathcal F}_{#1}} 

\newcommand{\remove}[1]{}
\allowdisplaybreaks

\begin{document}

\begin{frontmatter}
\title{On Covering a Solid Sphere with Concentric Spheres in~${\mathbb Z}^3$\tnoteref{ltitle}}
\tnotetext[ltitle]{A preliminary version of this work has appeared in ICAA'14 \cite{bera_14}.}

\author[isi]{Sahadev Bera}
\ead{sahadevbera@gmail.com}
\author[iit]{Partha Bhowmick\corref{corr}}
\ead{pb@cse.iitkgp.ernet.in, bhowmick@gmail.com}
\author[isi]{Bhargab B. Bhattacharya}
\ead{bhargab@isical.ac.in}

\cortext[corr]{Author for correspondence.}
\address[isi]{Advanced Computing and Microelectronics Unit, Indian Statistical Institute, Kolkata, India}
\address[iit]{Computer Science and Engineering Department, Indian Institute Technology, Kharagpur, India\bigskip\\
{\em September 05, 2014}}

\begin{abstract}
We show that a digital sphere, constructed by the circular sweep of a digital semicircle (generatrix) around
its diameter, consists of some holes (absentee-voxels), which appear on its spherical surface of revolution.
This incompleteness calls for a proper characterization of the absentee-voxels whose restoration will yield a
complete spherical surface without any holes. In this paper, we present a characterization of such
absentee-voxels using certain techniques of digital geometry and show that their count varies quadratically
with the radius of the semicircular generatrix. Next, we design an algorithm to fill these absentee-voxels so
as to generate a spherical surface of revolution, which is more realistic from the viewpoint of visual
perception. We further show that covering a solid sphere by a set of complete spheres also results in an
asymptotically larger count of absentees, which is cubic in the radius of the sphere. The characterization and
generation of complete solid spheres without any holes can also be accomplished in a similar fashion. We
furnish test results to substantiate our theoretical findings.
\end{abstract}


\end{frontmatter}


\hyphenation{compu-ta-tions compu-tation-ally corres-pond corres-pond-ing digital locate locat-ing loca-tion  
middle order parameter parameter-ization realistic revolu-tion rotate 
rota-tion rotat-ing surface theorem}

\section{Introduction}

Over the last two decades, the studies on geometric primitives in 2D and 3D digital space
have gained much momentum because of their numerous applications in computer graphics, image processing, 
and computer vision.
Apart from the characterization of straight lines and planes 
\cite{Brimkov_02,Brimkov_08b,Brimkov_07,Christ_12,Chun_09,Feschet_06,Fukshan_12,Kenmochi_08,Woo_08},
several theoretical work, mostly on digital spheres and hyperspheres, have appeared in the literature. 
A majority of them in 3D digital space are based on the extension of similar investigations on 
the characterization and generation of circles, rings, discs, and circular arcs in the 2D digital plane
\cite{andres_94,Andres_97,chan_95,davies_88,doros_84,haralick_74,nagy_04,pal_12,thomas_89,yuen_96}.
For various problems in science and engineering, discrete spheres are often required 
for simulation of experiments.
For example, in \cite{draine_94, zubko_10}, 
discrete spheres are used to test the accuracy of the {\em discrete dipole approximation}
(DDA) for computing scattering and absorption by isolated, homogeneous spheres, 
as well as by targets consisting of two contiguous spheres.
Hence, with the emergence of new paradigms, such as digital calculus \cite{nakamura_84}, 
digital geometry \cite{klette_04}, theory of words and numbers \cite{klette_04a,mignosi_91}, 
an appropriate characterization of a digital sphere is required to enrich our understanding of objects 
in 3D discrete space.

In this work, we address the problem of constructing a \emph{closed digital surface} defined by a set of 
points in $\zzz$ such that they optimally approximate a real sphere with integer radius.
Some prior work closely resemble our work, but they deal with spheres with a real value of radius.
For example, there is a multitude of papers in the literature, which discuss how to find the lattice points 
on or inside a real sphere of a given radius
\cite{Brimkov_08a,Heathbrown_99,Chamizo_07,Chamizo_12,Fomenko_02,Magyar_07,Ewell_00,Tsang_00}.
Some of them addresses the problem of finding a real sphere that passes through a given set of lattice points 
\cite{Maehara_10}.
They are closely related to the determination of lattice points on circles \cite{Cappell_07,honsberger_73}, 
ellipsoids \cite{Chamizo_09,Kuhleitner_00}, or on several types of surfaces of revolution
\cite{Chamizo_98}.

For hypersphere generation, characterization of a discrete analytical hypersphere has been done 
in \cite{Andres_97} to develop an algorithm, which is an extension of the algorithm for generating 
discrete analytical circles.
The algorithm is, however, quite expensive owing to complex operations in the real space.
An extension of the idea used in \cite{Andres_97} has been done in \cite{Fiorio_06b} based on a non-constant 
thickness function \cite{Fiorio_06a}, but no algorithm for generation of a digital sphere or hypersphere 
has been proposed.
Recently, analytical descriptions of various classes of digital circles,
spheres, and some cases of hyperspheres in a morphological framework
have been proposed in \cite{Toutant_13}.
Very recently, the notion of {\em discrete spherical geodesic path} between two voxels lying on 
a discrete sphere has been introduced in \cite{biswas_14},
and a number-theoretic algorithm has been proposed for construction of such paths in optimal time.

In \cite{Montani_90}, an algorithm for digitization of a real sphere with integer radius 
has been proposed.
It constructs the sphere as a sequence of contiguous digital circles by using Bresenham's circle 
drawing algorithm.
Such an approach fails to ensure the completeness of the generated digital sphere, since
it gives rise to absentee (missing) voxels, as shown in this paper.
The digital sphere generated by our algorithm, on the contrary, does not have any absentee-voxel,
since it fixes these absentees based on a digital-geometric characterization.

The work proposed in this paper aims to locate and fill the absentee-voxels 
(3D points with integer coordinates) on a digital spherical surface of revolution.
Covering such a surface by coaxial digital circles (with integer radius and
integer center) in $\zzz$ cannot produce the desired completeness of the surface 
owing to absentee-voxels.
Interestingly, the occurrence of absentees in such a cover is possibly a lesser fact.
The greater fact is that the absentees occur in multitude---an observation that motivates the requirement of their 
proper characterization, which subsequently aids~in designing a proven algorithm to generate a 
complete spherical surface in~$\mathbb Z^3$.

We have organized the rest of the paper as follows.
In Sec.~\ref{sec:preli}, we introduce few definitions and important properties related with 
digital circles, digital discs, and digital spheres considered in our work.
In Sec.~\ref{sec:sphere_abs}, we derive the necessary and sufficient condition for a voxel to be an absentee
in a sphere of revolution.
We also prove that the absentee count while covering a digital sphere of radius $r$ by 
coaxial digital circles---generated by the circular sweep of a digitally circular arc of 
radius $r$ (digital generatrix)---varies quadratically with~$r$.
In Sec.~\ref{ss:familysphere}, we characterize the absentee family, and use it 
in Sec.~\ref{ss:fixing_abs} for fixing the absentees in our proposed algorithm 
for generating a complete (i.e., absentee-free) sphere of revolution.
In Sec.~\ref{sec:solid}, we discuss further about the absentees in covering a solid sphere 
by union of complete spheres.
We show here that these absentees are of two kinds: {\em absentee lines} and {\em absentee circles}.
We derive their characterization in Sec.~\ref{ss:familysolidsphere}.
We use this characterization in Sec.~\ref{ss:count_solid} to show that the absentee counts corresponding to 
absentee lines and absentee circles are $\Theta(r^{5/2})$ and $\Theta(r^3)$ respectively.
The algorithm for fixing these absentees while generating a complete solid sphere 
is given in Sec.~\ref{ss:fixing_abs_solid}.
Finally, in Sec.~\ref{sec:results}, we present some test results to substantiate our 
theoretical findings.

\section{Preliminaries}
\label{sec:preli}

There exist several definitions of digital circles (and discs, spheres, etc.) in the literature, 
depending on whether the radius and the center coordinates are real or integer values.
Irrespective of these definitions, a digital circle (sphere) is essentially a set of points with integer coordinates, 
which are called {\em digital points} or {\em pixels} ({\em voxels})~\cite{klette_04}.
In this paper, we consider the {\em grid intersection digitization}~\cite{klette_04,stell_07} of a real
circle with integer radius and having center with integer coordinates.
Such a digitization produces a digital circle, which can be generated by the well-known {\em midpoint circle algorithm}
or the {\em Bresenham circle algorithm}~\cite{foley_93}, and its definition is as follows.


\begin{definition}[Digital circle]A digital circle with radius $r\inzp$ and center $o(0,0)$ 
is given by 
$\dc{r}=\Big\{(i,j)\in\zz:\begin{array}{c} \big|\max(|i|,|j|) -
\sqrt{r^2-(\min(|i|,|j|))^2}\big| <\frac12\end{array}\Big\}.$ 
\label{def:dcir-gen}\end{definition}

The points in $\dc{r}$ are connected in 8-neighborhood.
The points defining its {\em interior} are connected in 4-neighborhood, and hence separated by $\dc{r}$ 
from its {\em exterior} points, which are also connected in 4-neighborhood \cite{klette_04}.

All the results in this paper are valid for any non-negative integer radius and any center with integer coordinates.
So, for sake of simplicity, henceforth we consider the center as $o$ and use the notation 
$\dc{r}$ instead of $\dc{o,r}$, where $r\in{\mathbb Z}^+\cup\{0\}$.
We specify it explicitly when the center is not~$o$.

\begin{figure}[!t]\center
\includegraphics[width=\textwidth]{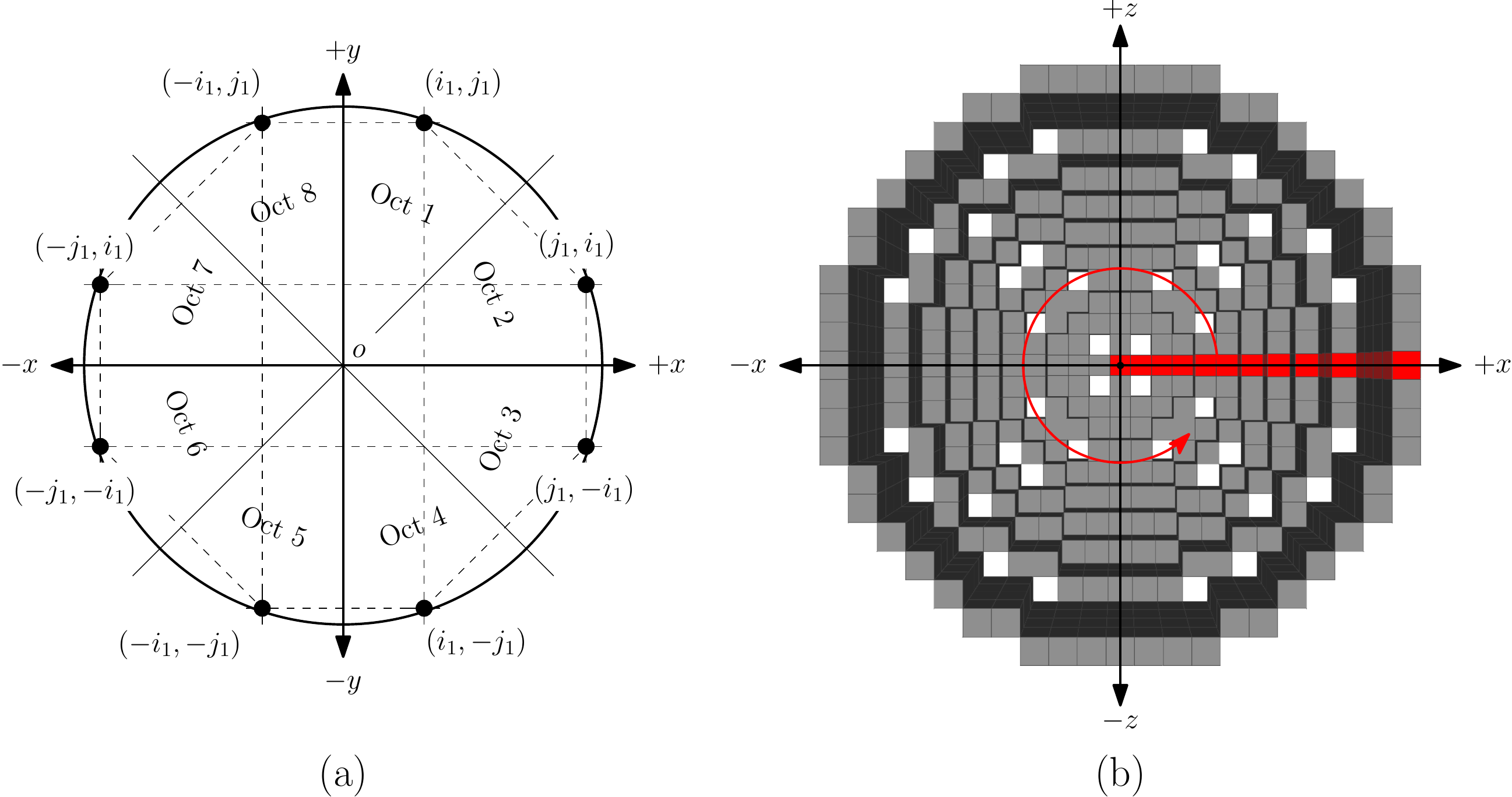}
\caption{(a)\,8-symmetric points $\{(i,j):\{|i|\}\cup\{|j|\}=\{i_1,j_1\}\}$ in eight respective octants of 
a digital circle $\dc{r}$.
(b)\,$\hr{r}$ for $r=10$, with the $+y$~axis pointing inwards w.r.t. the plane
of the paper.}
\label{fig:oct}
\end{figure}

A real point or a pixel $(x,y)$ is said to be lying in Octant~1 if and only if\, $0\leqslant x\leqslant y$
(Figure~\ref{fig:oct}(a)).
We use the notation $\dco{1}{r}$ to denote Octant~1 of 
$\dc{r}$, and ${\mathbb Z}_1^2$ to denote all points in Octant~1 of $\mathbb Z^2$.


\begin{definition}[Digital disc]
A digital disc of radius $r$ consists of all digital points in $\dc{r}$ and its interior, 
and is given by \[\dd{r}=\Big\{(i,j_c)\in\zz: \ 0\leq i\cdot i_c\leq i_c^2 \ \wedge
\left|\max(|i_c|,|j_c|) - \sqrt{r^2-(\min(|i_c|,|j_c|))^2}\right| <\frac12\Big\}.\]
\label{def:ddisc-gen}\end{definition}

Note that in Def.~\ref{def:ddisc-gen}, the condition $0\leq i\cdot i_c\leq i_c^2$ relates a disc pixel
$(i,j_c)$ to a circle pixel $(i_c,j_c)$, as $0\leq i_c$ implies $0\leq i\leq i_c$ and $i_c\leq 0$
implies $i_c\leq i\leq 0$.
If we consider the union of all digital circles centered at $o$ and radius in $\{0,1,2,\ldots,r\}$, then 
the resultant set $\du{r}:=\bigcup\limits_{s=1}^r\dc{s}$
is not identical with the digital disc of radius $r$.
The set $\du{r}$ contains absentee-pixels, as defined below.


\begin{definition}[Disc absentee]
A pixel $p$ is a disc absentee if and only if there exists some $r'\in\{1,2,\ldots,r\}$ such that 
$p$ is a point in the interior of $\dc{r'}$ and in the exterior of $\dc{r'-1}$.
\end{definition}

The above definition implies that if $p$ is any disc absentee, then $p$ does not belong 
to any digital circle, i.e., $p\in\dd{r}$ and $p\not\in\du{r}$.
Hence, the set of disc absentees is given by 
$\ad{r}=\dd{r}\sm \du{r}$.
The above definition of disc absentee is used in the following definitions
related to {\em spherical surfaces of revolution} in $\zzz$.
However, henceforth we do not use the term ``{\em of revolution}'' for sake of simplicity.
We also drop the term ``{\em digital}'' from any digital surface in $\zzz$.

Let $\dco{1,2}{r}$ denote the first quadrant (comprising the first and the second octants) of $\dc{r}$,
which is used as the {\em generatrix}.
When we rotate $\dco{1,2}{r}$ about $y$-axis through $360^o$, 
we get a stack (sequence) of circles representing a {\em hemisphere}, 
namely $\hr{r}:= \bigcup\limits_{(i,j)\in\dco{1,2}{r}}\dc{c,i}$,
where $c=(0,j,0)$ denotes the center of $\dc{c,i}$, as shown in Figure~\ref{fig:oct}(b).
Each circle $\dc{c,i}$ in this stack is generated by rotating a pixel $(i,j)\in\dco{1,2}{r}$
about $y$-axis.
The previous circle in the stack is either $\dc{c',i-1}$ or $\dc{c'',i}$,
where $c'=(0,j',0)$ with $j'\in\{j,j+1\}$, and $c''=(0,j+1,0)$.
There is no absentee between $\dc{c,i}$ and $\dc{c'',i}$, as they have the same radius.
But as the radii of $\dc{c,i}$ and $\dc{c',i-1}$ differ by unity,
there would be absentees (Definition~\ref{def:sphere_abs}) between them in $\hr{r}$.
Each such absentee $p$ would lie on the plane of $\dc{c,i}$ in the exterior of $\dc{c,i-1}$,
since $p$ did not appear in the part of $\hr{r}$ constructed up~to $\dc{c',i-1}$
and appeared only after constructing $\dc{c,i}$.
Hence, we have the following definition.

\begin{definition}[Sphere absentee]
A voxel $p$ is a sphere absentee lying on the plane $y=j$ if and only if there exist two 
consecutive points $(i,j)$ and $(i-1,j')$ in $\dco{1,2}{r}$, $j'\in\{j,j+1\}$, such that 
$p$ lies in the interior of $\dc{c,i}$ and in the exterior of $\dc{c,i-1}$, where $c=(0,j,0)$.
\label{def:sphere_abs}
\end{definition}

On inclusion of the sphere absentees (lying above $zx$-plane) with $\hr{r}$, we get the 
{\em complete hemisphere}, namely $\he{r}$.
On taking $\hr{r}$ and its reflection on $zx$-plane, we get the {\em sphere}, namely $\sr{r}$.
Similarly, the union of $\he{r}$ and its reflection on $zx$-plane gives the 
{\em complete sphere} in $\zzz$.
Let $\as{r}$ be the set of sphere absentees.
The number of points in $\as{r}$ is double the absentee count in $\hr{r}$.
We have the following definitions on spheres and their absentees.

\begin{definition}[Complete sphere]
A complete (hollow) sphere of radius $r$ is given by $\sp{r}=\sr{r}\cup\as{r}$.
\label{def:sphere} 
\end{definition}

\begin{definition}[Complete solid sphere]
A complete solid sphere $\ss{r}$ of radius $r$ is given by the union of $\sp{r}$ and 
voxels lying inside $\sp{r}$.
\label{def:sphere_solid} 
\end{definition}

\begin{definition}[Solid sphere absentee]
A voxel $p$ is a solid sphere absentee if and only if $p\in\ss{r}\sm\sa{r}$,
where $\sa{r}=\bigcup\limits_{r'=0}^r\sp{r'}$.
\label{def:sphere_abs_solid}
\end{definition}

\subsection{Previous Results}

We need the following results from \cite{bera_13} to count and fix the absentees
in the surface of revolution.

\begin{theorem}\label{thm:N_r}
The total count of disc absentees lying in $\du{r}$ is given by
\[|\ad{r}|=8\sum\limits_{k=0}^{m_r-1} |\ap{k}{r}|,\]
where $|\ap{k}{r}|=\left\lceil\sqrt{(2k+1)r-k(k+1)}\right\rceil-
\left\lceil{2k+1+\frac12\sqrt{(8k^2+4k+1)}}\right\rceil$\\
and $m_r=r-\left\lceil{r/\sqrt2}\right\rceil+1$.
\end{theorem}

\begin{theorem}
\label{thm:absBoundDisc}
$|\ad{r}|=\Theta(r^2)$.
\end{theorem}

\section{Absentees in a Digital Sphere}
\label{sec:sphere_abs}

As mentioned earlier in Sec.~\ref{sec:preli}, the hemisphere and the sphere 
have absentee-voxels, which can be characterized based on their unique correspondence 
with the absentee-pixels of $\du{r}$.
To establish this correspondence, we consider two consecutive pixels $p_i(x_i,y_i)$ and $p_{i+1}(x_{i+1},y_{i+1})$ 
of the generating curve $\dco{1,2}{r}$ corresponding to $\hr{r}$.
We have three possible cases as follows.
\begin{enumerate}
\item $(x_{i+1},y_{i+1})=(x_i+1,y_i)$ (Octant~1)
\item $(x_{i+1},y_{i+1})=(x_i+1,y_i-1)$ (Octant~1 or Octant~2)
\item $(x_{i+1},y_{i+1})=(x_i,y_i-1)$ (Octant~2)
\end{enumerate}

For Case~1, we get two concentric circles of radii differing by unity and lying on the same plane; 
the radii of the circles corresponding to $p_i$ and $p_{i+1}$ are $x_i$ and $(x_{i+1}=)x_i+1$.
Hence, for Case~1, the absentee-voxels between two consecutive circles easily correspond to the absentee-pixels
between $\dc{o,x_i}$ and $\dc{o,x_i+1}$.

For Case~2, the circle generated by $p_{i+1}$ has radius $x_{i+1}=x_i+1$ and its plane lies one voxel apart w.r.t. the plane of the circle generated by $p_i$.
Hence, if these two are circles are projected on $zx$-plane, then the absentee-pixels lying between the
projected circles have a correspondence with the absentee-voxels between the original circles.

For Case~3, we do not have an absentee, as the circles generated by $p_i$ and $p_{i+1}$ 
have the same radius ($x_i=x_{i+1}$).

Hence, the count of absentee-voxels in $\hr{r}$ is same as the 
count of absentee-pixels in $\du{r}$. 
However, it may be noted that the count of voxels present in $\hr{r}$
would be greater than the count of pixels present in $\du{r}$,
since for each circle of a particular radius $r'\in\{0,1,2,\ldots,r\}$ 
in $\du{r}$, there would be one or more circles 
of radius $r'$ (in succession) in $\hr{r}$.
We have the lemma on the correspondence of absentee count in 
$\hr{r}$ with that in $\du{r}$.

\begin{lemma}
\label{lem:one-one}
If $p(i,j,k)$ is an absentee-voxel in $\hr{r}$, 
then the pixel $(i,k)$ obtained by projecting $p$ on $xz$-plane is an absentee-pixel 
in $\du{r}$.
\end{lemma}

\begin{figure}[!t]\center
\begin{tabular}{@{}c@{ }c@{ }c@{}}
\includegraphics[scale=.38, page=1]{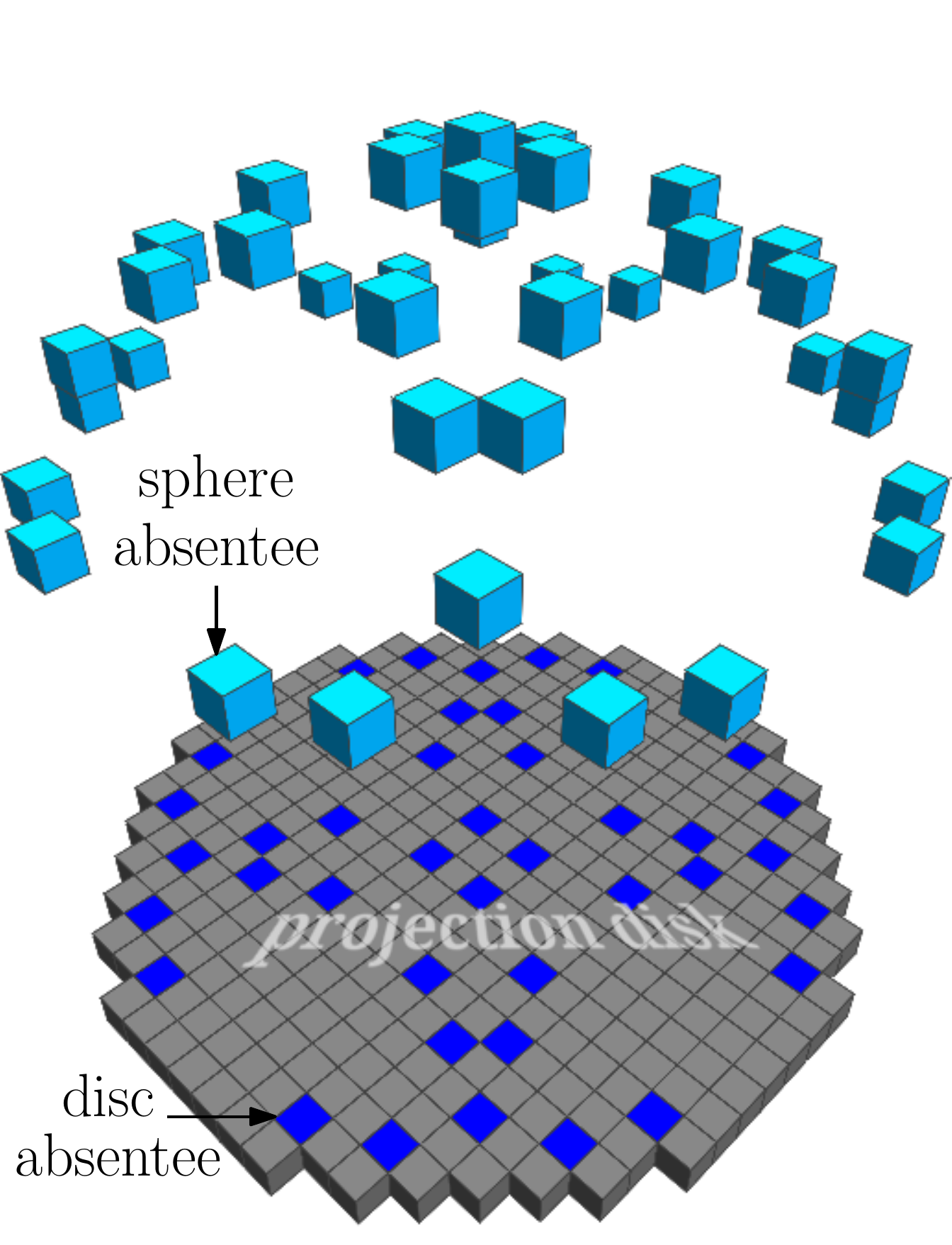}&
\includegraphics[scale=.38, page=2]{hsp-r10ab.pdf}&
\includegraphics[scale=.38, page=3]{hsp-r10ab.pdf}\\
(a)&(b)&(c)
\end{tabular}
\caption{(a) One-to-one correspondence for $r=10$ between absentee-voxels (shown in red) 
in $\hr{r}$ and absentee-pixels (shown in blue) in $\du{r}$. 
(b)\,Hemisphere of $r=10$ after fixing absentees.
(c)\,Parabolic surfaces of translation, produced by translating $\sup(\ipc{1})$ and $\sup(\spc{1})$,
$h=0,1,2$, along $y$-axis.} 
\label{fig:ab-r10}
\end{figure}

The above one-to-one correspondence between the absentees in the hemispherical surface for 
radius $r=10$ and the absentees related to the disc of radius $r=10$ is shown 
in Figure~\ref{fig:ab-r10}.
This one-to-one correspondence between the absentee set in $\hr{r}$
and that in $\du{r}$ leads to the following theorem.

\begin{theorem}
\label{thm:absBoundHemi}
The total count of absentee-voxels in $\hr{r}$ 
is $|\ad{r}|=\Theta (r^2)$.
\end{theorem}

\begin{proof}
Follows from Lemma~\ref{lem:one-one} and Theorem~\ref{thm:absBoundDisc}.
\end{proof}

On taking the reflection of $\he{r}$ about the $zx$-plane, 
we get the complementary hemisphere, namely $\hf{r}$.
The set $\he{r}\cup\hf{r}$ is the sphere, $\sp{r}$, corresponding to
which we get double the count of absentee-voxels compared to that in $\he{r}$.
Hence, we have the following theorem.

\begin{theorem}\label{thm:absBoundSphere}
The total count of absentee-voxels lying on $\sr{r}$ is given by
\vspace*{-4mm}
\[|\as{r}|=2|\ad{r}|
=16\sum\limits_{k=0}^{m_r-1} |\ad{r}|
=\Theta (r^2).\]
\vspace*{-5mm}
\end{theorem}
\begin{proof}
Follows from Theorem~\ref{thm:N_r} and Theorem~\ref{thm:absBoundHemi}.
\end{proof}


\subsection{Characterizing the Absentee Family}
\label{ss:familysphere}

We use the following lemmas from \cite{bera_13,bhow_08b} 
for deriving the necessary and sufficient conditions to decide 
whether a given voxel is an absentee or not.

\begin{lemma}[circle pixel \cite{bhow_08b}]\label{lem:I_k}
The squares of abscissae of the pixels with $z=k$ in $\dco{1}{r'}$ 
drawn on $zx$-plane lie in the interval $I_{r'-k}^{(r')}:=\left[u_{r'-k}^{(r')}, v_{r'-k}^{(r')}\right)$, where
$u_{r'-k}^{(r')}=r'^2-k^2-k$ and $v_{r'-k}^{(r')}=r'^2-k^2+k$.
\end{lemma}

\begin{lemma}[absentee \cite{bera_13}]\label{lem:J_r}
A point $(i,0,k)\in\zzz$ is an absentee on the $zx$-plane if and only if $i^2$ lies in the integer interval
$J_{r'-k}^{(r')}:=\left[v_{r'-k}^{(r')},u_{r'+1-k}^{(r'+1)}\right)$ for some
$r'\inzp$.
\end{lemma}

We have now the following theorem on the necessity and sufficiency for an absentee-voxel
in $\sr{r}$.

\begin{theorem}\label{thm:abs_vox_cond}
A voxel $p(i,j,k)$ is an absentee if and only if $i^2\in J_{r'-k}^{(r')}$ for some $r'\inzp$
and $r'^2\in I_{r-j}^{(r)}$.
\end{theorem}

\begin{proof}
Lemma~\ref{lem:one-one} implies that when $p(i,j,k)$ is an absentee-voxel in 
$\hr{r}$, 
then its projection pixel $p'(i,k)$ on $zx$-plane is absentee-pixel in $\du{r}$.
Hence, by Lemma~\ref{lem:J_r}, $i^2$ lies in $J_{r'-k}^{(r')}$ for some $r'\inzp$.
What now remains to check is the condition for $y$-coordinate of $p$.
Observe that there exists a circle $\dc{c,r'}$ centered at $c=(0,j,0)$ 
on the hemisphere such that the the projection of $\dc{c,r'}$ on $zx$-plane 
is the circle of radius $r'$ in $\du{r}$.
Again $p(i,j,k)$ and $\dc{c,r'}$ lie on the same plane, i.e., $y=j$.
Hence, the pixel $(r',j)$ must lie on the generating circular arc, $\dco{1,2}{r}$,
and so by Lemma~\ref{lem:I_k}, we have $r'^2\in I_{r-j}^{(r)}$.

Conversely, if $i^2\in J_{r'-k}^{(r')}$, then $p\not\in\hr{r}$;
and if $r'^2\in I_{r-j}^{(r)}$ for some $r'\in\{0,1,2,\ldots,r\}$, then 
$p\in\he{r}$, wherefore $p$ is an absentee.
\end{proof}

An example of absentee-voxel is $(2,9,4)$ in hemisphere of radius $r=10$ (Figure~\ref{fig:ab-r10}), 
since for $k=4$, we have $r'=4$ for which $v_{r'-k}^{(r')}=r'^2-k^2+k=16-16+4=4$,
$u_{r'+1-k}^{(r'+1)}=(r'+1)^2-k^2-k=25-16-4=5$, thus giving $J_{0}^{(4)}=[4,5)=[4,4]$ in which lies
the square number $4=i^2$ and $u_{r-j}^{(r)}=r^2-j^2-j=10^2-9^2-9=10$,
$v_{r-j}^{(r)}=r^2-j^2+j=10^2-9^2+9=28$, thus giving $I_{r-j}^{(r)}=[10,28)$ which contains
$r'^2=16$.

On the contrary, $(3,9,4)$ is not an absentee-voxel, as for $k=4$, there is no such $r'$ for which
$J_{r'-4}^{(r')}$ contains $3^2$; in fact, for $k=4$, we get the interval
$I_{5-4}^{(5)}=[5^2-4^2-4,5^2-4^4+4)=[5,12]$ with $r'=5$, which contains $3^2$, thereby making
$(3,9,4)$ a point on hemisphere of radius $r=10$ at the plane $y=9$.


To characterize the absentees as a whole, we use Lemma~\ref{lem:J_r} for the expanded form of 
(the lower and the upper limits of) $J_{r'-k}^{(r')}$.
We replace $r'$ by $k+h$ and $r'+1$ by $k+(h+1)$, where the {\em $h(\geq0)$th run of pixels} in 
$\dco{1}{r'}$ drawn on $zx$-plane has $z=k$ \cite{bhow_08b}.
Thus,
\begin{equation}\begin{array}{l}
v_{r'-k}^{(r')}=(2h+1)k+h^2,\medskip\\
u_{r'+1-k}^{(r'+1)}=(2h+1)k+(h+1)^2.\end{array}
\label{eqn:uv_parab}\end{equation}
Hence, if $p(i,0,k)$ is a point in Octant~1 and lies on $h^{th}$ run of $\dco{1}{r'}$, 
then
\begin{equation}
i^2<(2h+1)k+h^2;
\label{eqn:ij1_parab}\end{equation}
and if $p(i,0,k)$ is a point in Octant~1 and lies left of the $(h+1)^{th}$ run of $\dco{1}{r'+1}$,
then
\begin{equation}
i^2<(2h+1)k+(h+1)^2.
\label{eqn:ij2_parab}\end{equation}
Equations~\ref{eqn:ij1_parab} and \ref{eqn:ij2_parab} correspond to two parabolic regions in 
the real ($zx$-)plane
on replacing $i$ and $k$ by $x$ and $z$, respectively, $h$ being considered as a constant.
These {\em open parabolic regions} are given by
\begin{equation}\begin{array}{l}
\ipc{1}: x^2<(2h+1)z+h^2,\medskip\\
\spc{1} : x^2<(2h+1)z+(h+1)^2.\end{array}
\label{eqn:PuPv}\end{equation}

The respective suprema of these two regions are given by two parabolas, namely 
$\sup(\ipc{1}): x^2=(2h+1)z+h^2$ and $\sup(\spc{1}): x^2=(2h+1)z+(h+1)^2$.
In 3D space, these two suprema correspond to two {\em parabolic surfaces of translation},
produced by translating $\sup(\ipc{1})$ and $\sup(\spc{1})$
along $y$-axis, as shown in Figure~\ref{fig:ab-r10}c.
Evidently, the absentees of $\hr{r}$ in Octant~1 and Octant~8
lie in the {\em half-open 3D parabolic region} given by $P_h:=\spc{1}\sm\ipc{1}$
for a given pair of $k$ and $h$, i.e., for a given $(r',k)$-pair.
The family of all the half-open 3D parabolic regions, $P_0, P_1, P_2, \ldots$, 
thus contains all the absentees $\hr{r}$ in Octant~1 and Octant~8, 
as stated in the following theorem.

\begin{theorem}\label{thm:F}
All the absentees of $\hr{r}$ in Octant~1 and Octant~8 lie in
\[{\mathcal F}:=\left\{P_h \cap {\mathbb Z}_1^3: h=0,1,2,\ldots\right\}.\]
\end{theorem}
\begin{proof}
Follows from Theorem~\ref{thm:abs_vox_cond} and Eqn.~\ref{eqn:PuPv}.
\end{proof}

\subsection{Fixing the Absentee-Voxels}
\label{ss:fixing_abs}

Algorithm~\ref{algo:AVH} (AVH) shows the steps for fixing the absentee-voxels corresponding to
the hemisphere $\hr{r}$ having radius $r$.
The generating curve, which is an input to this algorithm, is the circular arc, $\dco{1,2}{r}$.
This circular arc is a (ordered) sequence of points, $\{p_t(i_t,j_t,0)\in\zzz: t=1,2,\ldots,n_r\}$,
whose first point is $p_1(0,r,0)$ and last point is $p_{n_r}(r,0,0)$.
The point $p_{t+1}$ can have $i_{t+1}$ either same as $i_t$ of the previous point $p_t$
or greater than $i_t$ by unity.
For the former case, there is no absentee between the two circles generated by $p_t$ and $p_{t+1}$.
For the latter, the absentees are computed by invoking the procedure {\tt ACC}, as shown in 
Step~\ref{AVH:ACC_call} of Algorithm~\ref{algo:AVH}.

The procedure {\tt ACC} finds the absentee-voxels between two concentric circles, 
$\dct{y=j}{c,i}$ and $\dct{y=j}{c,i+1}$ of radii $i$ and $i+1$, each centered at $(0,j,0)$ 
on $y=j$ plane.
The set of all absentees between these two circles is denoted by $A$.
As an absentee lies just after the end of a {\em voxel-run} corresponding to the interval $I_{r-j}^{(r)}$ 
(Lemma~\ref{lem:I_k}), the procedure {\tt ACC} first computes the voxel-run in the plane $y=j$
(Steps~\ref{ACC:voxelRunStart}--\ref{ACC:voxelRunEnd}).
Then, in Step~\ref{ACC:voxelCheck}, it determines whether the next voxel is an absentee in Octant~1, 
using Lemma~\ref{lem:J_r}.
For each absentee-voxel in Octant~1, the absentees in all other octants are included 
in $A$, as shown in Step~\ref{ACC:absentees}.
Figure~\ref{fig:ab-r10}(a) shows the hemisphere for $r=10$, whose absentees (shown in red) have been
fixed by Algorithm~\ref{algo:AVH}.

\begin{figure}[!t]\center
\begin{minipage}{.95\linewidth}
\begin{algorithm}[H]
\caption{(AVH) Fixing absentee-voxels in the hemisphere}
\label{algo:AVH}
\KwIn{Generating circular arc, $\dco{1,2}{r}:=\{p_1,p_2,\ldots,p_{n_r}\}$}
\KwOut{Absentee-voxels in $\hr{r}$}\medskip

$\as{r}$ $\leftarrow \emptyset$\\
\For{$t=1,2,\ldots,n_r-1$}{
    \If{$i_{t+1}>i_t$}{
	$\as{r}\leftarrow\as{r}\cup{\tt ACC}(\dco{1,2}{r},t)$ \label{AVH:ACC_call}
    }
}
\Return{$\as{r}$}
\end{algorithm}
\end{minipage}\\
\begin{minipage}{.95\linewidth}
\begin{procedure}[H]
\caption{ACC($\dco{1,2}{r},t$)}
\label{proc:ACC}
$A\leftarrow\emptyset, r\leftarrow i_t$\\
{\bf int} $i\leftarrow 0, k\leftarrow r, s\leftarrow 0, w\leftarrow r-1$\\
{\bf int} $l\leftarrow 2w$\\
\While{$k\geq i$}{
    \Repeat{$s\leq w$}{                            \label{ACC:voxelRunStart}
	  $s\leftarrow s+2i+1$\\
	  $i\leftarrow i+1$}                       \label{ACC:voxelRunEnd}
    \If{$i^2\in J_{r-k}^{(r)}$ {\bf and} $k\geq i$ \label{ACC:voxelCheck}}
    {
        $A \leftarrow A \cup \{(i',j_t,k'): \{|i'|\}\cup\{|k'|\}=\{i,k\}\}$ 
        \label{ACC:absentees}
    }
    $w\leftarrow w+l$, $l\leftarrow l-2$, $k\leftarrow k-1$
}
\Return{$A$}
\end{procedure}
\end{minipage}
\end{figure}


\section{Absentees in a Solid Sphere}
\label{sec:solid}

As mentioned earlier in Sec.~\ref{sec:preli}, the absentee-voxels in a solid sphere $\sa{r}$ 
can also be characterized using the set of disc absentees, $\ad{r}$.
The set of voxels defining $\sa{r}$ is given by the union of the voxel sets corresponding to 
the complete spheres of radii $0,1,2,\ldots,r$ (Definition~\ref{def:sphere_abs_solid}).

\begin{figure}[!t]\center
\begin{tabular}{@{}c@{ }c@{ }c@{}}
\includegraphics[scale=.38, page=4]{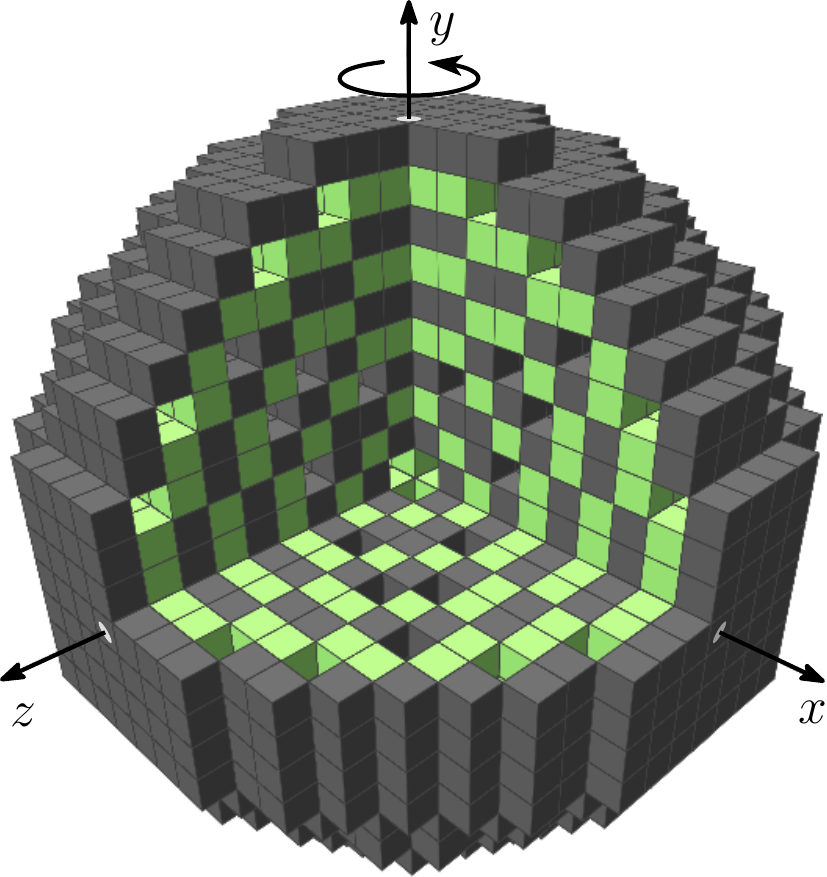}&
\includegraphics[scale=.38, page=5]{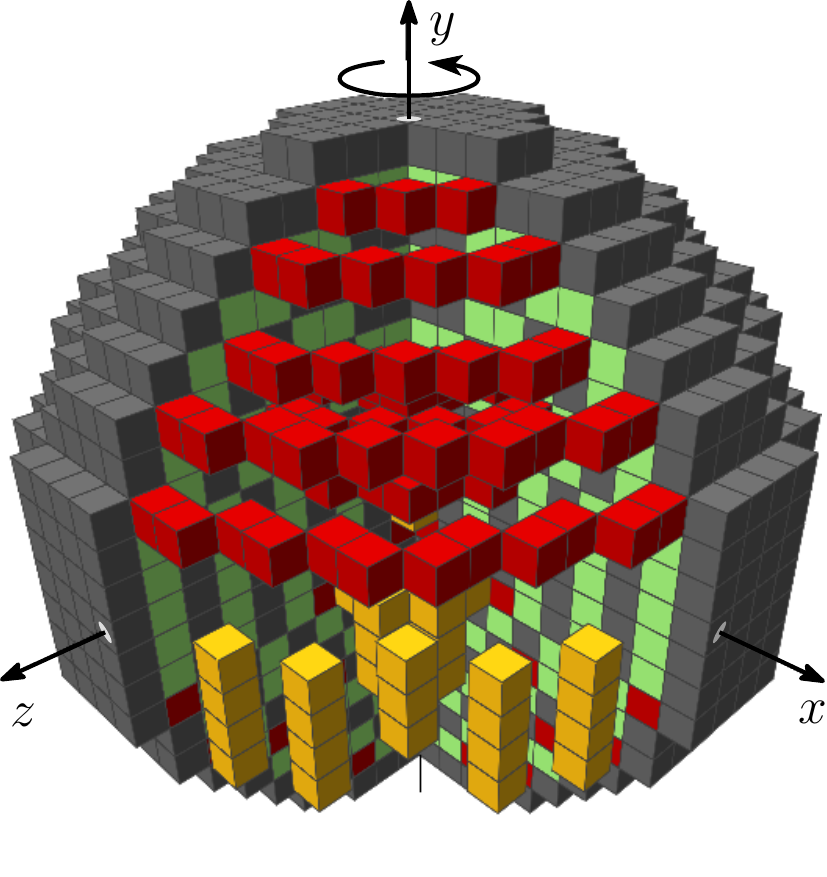}&
\includegraphics[scale=.38, page=6]{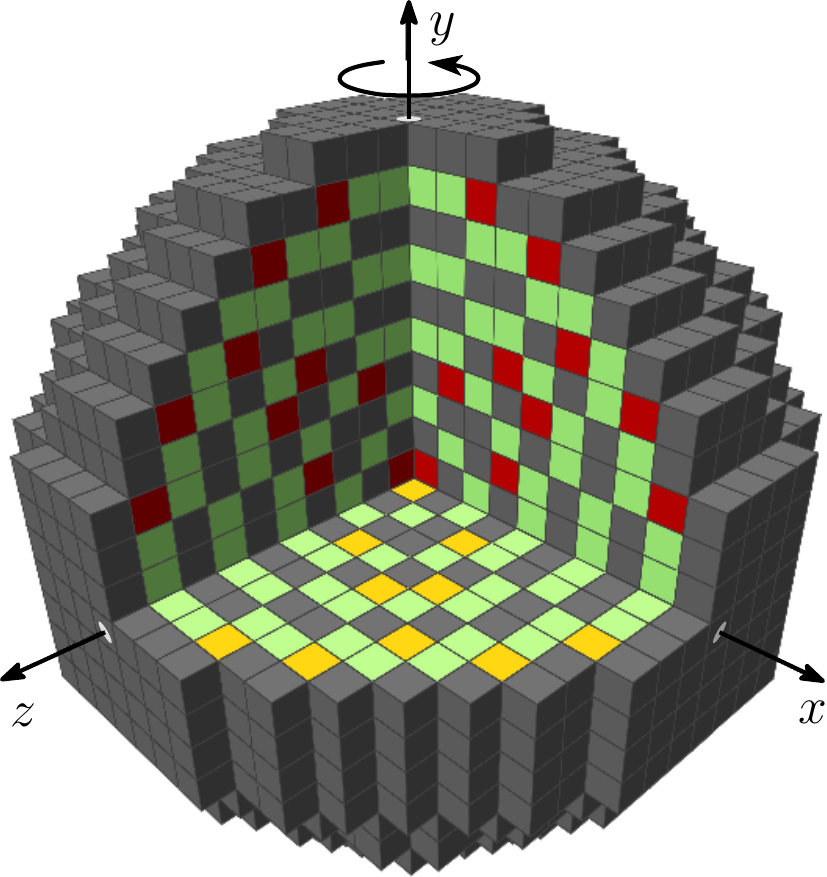}\\
(a)&(b)&(c)
\end{tabular}
\caption{Covering a solid sphere ($r=10$) by concentric complete spheres.
(a)\,Concentric complete spheres for $r=0,1,\ldots,10$.
(b)\,Absentee lines (yellow) in the lower hemisphere and 
     absentee circles (red) in the upper hemisphere.
(c)\,Complete solid sphere after fixing the absentee lines and absentee circles.}
\label{fig:solid_abs}
\end{figure}

To find the absentees in $\sa{r}$, we consider its lower (or upper) hemisphere, $\hrs{r}$.
Three-fourth of the upper hemisphere and the entire lower hemisphere of $\sa{r}$ are
shown in Figure~\ref{fig:solid_abs}a.
Observe that the set of voxels of $\hrs{r}$ lying in the 1st quadrant of $xy$-, $yz$-,
or $zx$-plane is given by the union of voxels comprising those arcs of the complete spheres
which lie in the 1st quadrant of the concerned plane.
Hence, the above set of voxels is same as the subset of $\du{r}$ lying in this quadrant,
or, the absentee-voxels in this quadrant are in one-to-one correspondence with 
the disc absentees in $\du{r}$ (Figure~\ref{fig:ab-r10}a).
The absentees in this quadrant are, however, characterized depending on the 
coordinate plane, as follows.

\begin{enumerate}
  \item[(AL)] For each absentee $p(i,0,k)$ in the 1st quadrant of $zx$-plane, 
there are absentees in $\hrs{r}$, which comprise an {\em absentee line},
given by $\al{i,k}=\{(i,j,k): j'<j\leq 0 \wedge (i,j',k)\in\hrs{r}\}$.
These absentee lines are shown in yellow in the lower hemisphere in Figure~\ref{fig:solid_abs}b.
  \item[(AC)] For each absentee $p(i,j,0)$ in the 1st quadrant of $xy$-plane, 
there are absentees in $\hrs{r}$, which comprise an {\em absentee circle},
given by $\ac{i,j}=\{(i',j,k'): (i',j,k')\in\dct{y=j}{c,i} \wedge c=(0,j,0)\}$.
These absentee circles, shown in red in the upper hemisphere in Figure~\ref{fig:solid_abs}b,
pass through the absentees in the 1st quadrant of $yz$-plane.
\end{enumerate}

\subsection{Characterizing the Absentee Family}
\label{ss:familysolidsphere}

We characterize here the family of absentee-voxels comprising the absentee lines 
that comprising the absentee circles.
For this, we need the following theorems on the necessity and sufficiency for an absentee 
belonging to an absentee line or an absentee circle in $\hrs{r}$.

\begin{theorem}\label{thm:abs_line}
A voxel $p(i,j,k)$ belongs to an absentee line if and only if $i^2\in J_{r'-k}^{(r')}$ 
for some $r'\inzp$ and $0\leq j\leq \lfloor{\sqrt{r'}}\rfloor+1$.
\end{theorem}

\begin{proof}
From (AL), if $p(i,j,k)$ belongs to an absentee line $\al{i,k}$, then 
$p(i,0,k)$ is a disc~absentee on $zx$-plane.  
Hence, by Lemma~\ref{lem:J_r}, $i^2\in J_{r'-k}^{(r')}$ for some $r'\inzp$.
Further, $\al{i,k}$ lies between two complete hemispheres, namely $\hrs{r'}$ and $\hrs{r'+1}$,
where $r'=\left\lfloor i^2+k^2-k \right\rfloor$. 
In~particular, $\al{i,k}$ lies between the surface of revolution generated by the topmost run
of (generatrix) $\dco{34}{r'+1}$ and that generated by the topmost run of $\dco{34}{r'}$.
Hence, the value of $j$ can be at~most the length of the topmost run of
$\dco{34}{r'+1}$, which is $\lfloor\sqrt{r'}\rfloor+1$ by Lemma~\ref{lem:I_k}.

Conversely, if $i^2\in J_{r'-k}^{(r')}$ for some $r'\inzp$, then
by Lemma~\ref{lem:J_r}, $p(i,0,k)$ is a disc~absentee on $zx$-plane. 
Hence, if $0\leq j\leq \lfloor{\sqrt{r'}}\rfloor+1$,
then from (AL), $(i,j,k)$ belongs to $\al{i,k}$ that lies between the surfaces of revolution 
generated by $\dco{34}{r'+1}$ and $\dco{34}{r'}$.~\end{proof}

\begin{theorem}\label{thm:abs_cir}
A voxel $p(i,j,k)$ belongs to an absentee circle if and only if $i^2\in I_{r'-k}^{(r')}$ 
for some $r'\inzp$ such that $r'^2\in J_{r''-j}^{(r'')}$ for some $r''\inzp$.
\end{theorem}

\begin{proof}
From (AC), if $p(i,j,k)$ belongs to an absentee circle, then
that absentee circle $\dc{r'}$ lies on the plane $y=j$,
where $(r',j)$ is a disc~absentee in $\du{r}$ in $xy$-plane.
Hence, by Lemma~\ref{lem:J_r}, $r'^2\in J_{r''-j}^{(r'')}$ for some $r''\inzp$.
Further, since $p\in\dc{r'}$, we get $i^2\in I_{r'-k}^{(r')}$ by Lemma~\ref{lem:I_k}.

Conversely, if $r'^2\in J_{r''-j}^{(r'')}$ for some $r''\inzp$,
then by Lemma~\ref{lem:J_r}, $(r',j)$ is a disc~absentee on $xy$-plane,
and so $\dc{r'}$ is an absentee circle.
Hence, by Lemma~\ref{lem:I_k}, if $i^2\in I_{r'-k}^{(r')}$, then $p\in\dc{r'}$,
or, $p$ belongs to an absentee circle.
\end{proof}


The absentee circles are characterized based on locations of their corresponding disc absentees
on (real) $xy$-plane.
The digital disc has eight octants.
All the absentee circles corresponding to the disc absentees in Octant~1 include the disc absentees
in Octant~8.
Reflection of these absentee circles simply gives all the absentee circles corresponding to 
the disc absentees in Octant~4 (and~5), and hence the characterization of 
absentee circles corresponding to Octant~4 is very much similar to that corresponding to Octant~1.
But the characterization of absentee circles corresponding to Octant~2 (and~7) 
is different and it can be used to obtain the characterization corresponding to Octant~3 (and~6) also.
Hence, following are two theorems on characterization of absentee circles---one for Octant~1
and another for Octant~2.

\begin{figure}[t]\center
\includegraphics[width=5in]{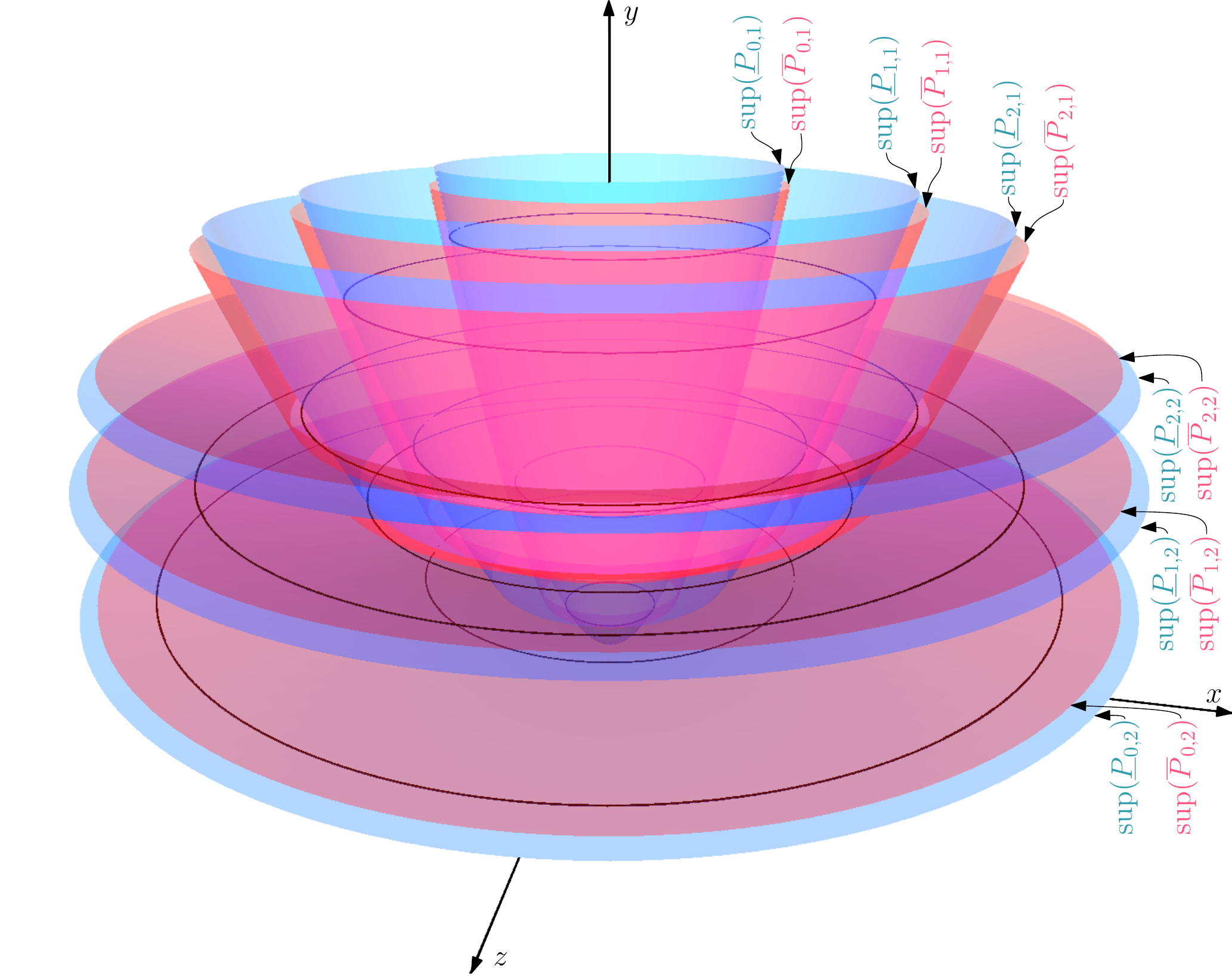}
\caption{Illustration of Theorem~\ref{thm:F_1} and Theorem~\ref{thm:F_2} for $r=10$.
The paraboloidal surfaces $\sup(\ipc{1})$ and $\sup(\ipc{2})$ are shown in blue,
and $\sup(\spc{1})$ and $\sup(\spc{2})$ in red; $h=0,1,2$ as $r=10$.
For clarity, the absentee circles lying in the open paraboloidal spaces are shown as real circles.} 
\label{fig:ab-r10solid-char}
\end{figure}

\begin{theorem}\label{thm:F_1}
All the absentee circles of $\sa{r}$ corresponding to Octant~1 lie in
\begin{equation}
\fa{1}:=\left\{\pv{h}{1} \cap {\mathbb Z}_1^3: h=0,1,2,\ldots\right\}
\label{eqn:F_1}\end{equation}
where, 
\begin{equation}
\pv{h}{1}=\sps{1}\sm\ips{1}
\label{eqn:pv_h_1}\end{equation}
such that
\begin{equation}\begin{array}{l}
\ips{1}:x^2+z^2<(2h+1)y+h^2,\medskip\\
\sps{1}:x^2+z^2<(2h+1)y+(h+1)^2.\end{array}
\label{eqn:ips_sps_1}\end{equation}
\end{theorem}
\begin{proof}
We use Theorem~\ref{thm:abs_cir} for the expanded forms of the lower and the upper limits of
$J_{r''-j}^{(r'')}$.
Using Lemma~\ref{lem:J_r} and replacing $r''-j$ by $h$, we get
\begin{equation}\begin{array}{l}
v_{r''-j}^{(r'')}=(2h+1)j+h^2,\medskip\\
u_{r''+1-j}^{(r''+1)}=(2h+1)j+(h+1)^2.\end{array}
\label{eqn:uv_parab_1}\end{equation}

Hence, if $p(i,j=r''-h,0)$ is a point on or inside $\dco{1}{r''}$
but strictly inside $\dco{1}{r''+1}$, then
\begin{equation}\begin{array}{l}
i^2<(2h+1)j+h^2,\medskip\\
i^2<(2h+1)j+(h+1)^2.\end{array}
\label{eqn:ij_parab}\end{equation}

Equation~\ref{eqn:ij_parab} corresponds to two {\em open parabolic regions} 
on the $xy$-plane, on replacing $i$ and $j$ by $x$ and $y$ respectively, 
$h$ being considered as a constant.
These open parabolic regions are given by
\begin{equation}\begin{array}{l}
\ipc{1}: x^2<(2h+1)y+h^2,\medskip\\
\spc{1}: x^2<(2h+1)y+(h+1)^2.\end{array}
\label{eqn:PuPv_so}\end{equation}

The respective suprema of these two regions are given by two parabolas, namely 
$\sup(\ipc{1}): x^2=(2h+1)y+h^2$ and $\sup(\spc{1}): x^2=(2h+1)y+(h+1)^2$.
On rotating $\sup(\ipc{1})$ and $\sup(\spc{1})$ about $y$-axis, 
we get two paraboloidal surfaces that enclose two {\em open paraboloidal spaces} given by
Eqn.~\ref{eqn:ips_sps_1}.
Evidently, the absentee circles corresponding to the disc absentees in Octant~1 and Octant~8 
lie in the {\em half-open paraboloidal space} $\pv{h}{1}$, given by Eqn.~\ref{eqn:pv_h_1},
for a given value of $h(=r''-j)$.
Hence, the family $\fa{1}$ of all these half-open paraboloidal spaces is given by Eqn.~\ref{eqn:F_1},
which contains all the aforesaid absentee circles.
An illustration is shown in Figure~\ref{fig:ab-r10solid-char} for $r=10$.
\end{proof}

\begin{theorem}\label{thm:F_2}
All the absentee circles of $\sa{r}$ corresponding to Octant~2 lie in
\begin{equation}
\fa{2}:=\left\{\pv{h}{2} \cap {\mathbb Z}_1^3: h=0,1,2,\ldots\right\}
\label{eqn:F_2}\end{equation}
where, 
\begin{equation}
\pv{h}{2}=\sps{2}\sm\ips{2}
\label{eqn:pv_h_2}\end{equation}
such that
\begin{equation}\begin{array}{l}
\ips{2}:y^2<(2h+1)\sqrt{x^2+z^2}+h^2,\medskip\\
\sps{2}:y^2<(2h+1)\sqrt{x^2+z^2}+(h+1)^2.\end{array}
\label{eqn:ips_sps_2}\end{equation}
\end{theorem}
\begin{proof}
In Octant~1, the parabolic regions containing the disc absentees have all their axes 
coinciding with $y$-axis.
On the contrary, in Octant~2, the axes of the parabolic regions containing the disc absentees 
all coincide with $x$-axis.
Hence, for each absentee $(i,j)$ in Octant~2, $j^2\in J_{r''-i}^{(r'')}$.
We use Theorem~\ref{thm:abs_cir} as before, and using Lemma~\ref{lem:J_r} and replacing $r''-i$ by $h$, 
we get
\begin{equation}\begin{array}{l}
v_{r''-i}^{(r'')}=(2h+1)i+h^2,\medskip\\
u_{r''+1-i}^{(r''+1)}=(2h+1)i+(h+1)^2.\end{array}
\label{eqn:uv_parab_2}\end{equation}

Hence, if $p(i=r''-h,j,0)$ is a point on or inside $\dco{1}{r''}$
but strictly inside $\dco{1}{r''+1}$, then
\begin{equation}\begin{array}{l}
j^2<(2h+1)i+h^2,\medskip\\
j^2<(2h+1)i+(h+1)^2.\end{array}
\label{eqn:ji_parab}\end{equation}

As explained in Theorem~\ref{thm:F_1}, Eqn.~\ref{eqn:ji_parab} corresponds to two 
{\em open parabolic regions} given by
\begin{equation}\begin{array}{l}
\ipc{2}: y^2<(2h+1)x+h^2,\medskip\\
\spc{2}: y^2<(2h+1)x+(h+1)^2.\end{array}
\label{eqn:ipc_spc_2}\end{equation}

The respective suprema of these two regions are two parabolas, namely 
$\sup(\ipc{2}): y^2=(2h+1)x+h^2$ and $\sup(\spc{2}): y^2=(2h+1)x+(h+1)^2$.
On rotating $\sup(\ipc{2})$ and $\sup(\spc{2})$ {\em again about $y$-axis}, 
we get two paraboloidal surfaces that enclose two {\em open paraboloidal spaces} given by
Eqn.~\ref{eqn:ips_sps_1}.
The absentee circles corresponding to the disc absentees in Octant~2
lie in the {\em half-open paraboloidal space} $\pv{h}{2}$,
given by Eqn.~\ref{eqn:pv_h_2}, for a given value of $h(=r''-i)$.
As in Theorem~\ref{thm:F_1}, the family $\fa{2}$, given by Eqn.~\ref{eqn:F_2},
contains all the absentee circles corresponding to Octant~2.
See Figure~\ref{fig:ab-r10solid-char} for an illustration with $r=10$.
\end{proof}

\subsection{Absentee Count}
\label{ss:count_solid}

We first have the following lemma on the count of absentee lines and that of absentee circles.
\begin{lemma}
The respective counts of absentee lines and absentee circles in $\hrs{r}$
are $|\ad{r}|$ and $\frac{1}{4}|\ad{r}|$.
\label{lem:ab_lc}
\end{lemma}

Using Lemma~\ref{lem:ab_lc}, we derive the count of absentee-voxels in $\sa{r}$,
as stated in the following theorem.

\begin{theorem}\label{thm:absBoundSolidSphere}
The count of all absentee-voxels in $\sa{r}$ is $\Theta (r^3)$.
\end{theorem}
\begin{proof}
We first count the absentee-voxels comprising the absentee lines in $\hrs{r}$.
From Theorem~\ref{thm:abs_line}, 
the count of voxels in $\al{i,k}$ is given by the length of the topmost run of
$\dco{34}{r'+1}$, which is $\lfloor\sqrt{r'}\rfloor+1$ by Lemma~\ref{lem:I_k}.

From Theorem~\ref{thm:absBoundDisc}, the count of disc absentees in $\du{r'}$ is $\Theta(r'^2)$,
which implies that the count of disc absentees between $\dc{r'}$ and $\dc{r'+1}$ is $\Theta(r')$.
Hence, the count of absentee-voxels comprising all the absentee lines in $\hrs{r}$ is
\begin{equation}
\sum\limits_{r'=1}^{r} \Theta(r') \Theta(\sqrt{r'}) 
= \sum\limits_{r'=1}^{r} \Theta\left({r'}^{\,3/2}\right) 
= \Theta\left(\sum\limits_{r'=1}^{r}{r'}^{\,3/2}\right)
= \Theta\left(r^{5/2}\right),
\label{eq:lineabs}
\end{equation}
since $\sum\limits_{r'=1}^{r}{r'}^{\,3/2} > \sum\limits_{r'=\lfloor r/2 \rfloor}^{r}{r'}^{\,3/2}$, 
or, $\sum\limits_{r'=1}^{r}{r'}^{\,3/2}= \Omega(r^{5/2})$,\\
and $\sum\limits_{r'=1}^{r}{r'}^{\,3/2} < \sum\limits_{r'=1}^{r}(r)^{3/2}$,
or, $\sum\limits_{r'=1}^{r}{r'}^{\,3/2}= O(r^{5/2})$.

Now we count the absentee-voxels comprising the absentee circles in $\hrs{r}$.
From (AC), each absentee $p(i,j,0)$ corresponds to an absentee circle $\ac{i,j}$,
which has radius~$i$ and lies on the plane $y=j$.
Its symmetric absentee $p'(j,i,0)$ corresponds to another absentee circle $\ac{j,i}$,
which has radius~$j$ and lies on the plane $y=i$.
Voxel count of these two absentee circles is $\Theta(i)+\Theta(j)=\Theta(i+j)$, 
which is asymptotically same as the voxel count of $\ac{r',r',0}$, where $r'=i+j$.
As explained above, the count of disc absentees between $\dc{r'}$ and $\dc{r'+1}$ is $\Theta(r')$,
and for each of these disc absentees, $i+j=\Theta(r')$.
Hence, the count of absentee-voxels comprising all the absentee circles in $\hrs{r}$ is
\begin{equation}
\sum\limits_{r'=1}^{r} \Theta(r') \Theta(r') 
= \sum\limits_{r'=1}^{r} \Theta\left((r')^2\right) 
= \Theta\left(r^3\right).
\label{eq:circleabs}
\end{equation}
On doubling the absentee count as obtained above for $\hrs{r}$,
we get the count of all absentees in $\sa{r}$ as $\Theta (r^3)$.
\end{proof}

\subsection{Fixing the Absentee Lines and Circles}
\label{ss:fixing_abs_solid}

Algorithm~\ref{algo:AVS} (AVS) shows the steps for fixing the absentee lines and absentee circles 
corresponding to the solid sphere $\sa{r}$ having radius $r$.
For each disc absentee in Octant~1 on $zx$-plane, 
there are four or eight absentee lines, which are computed by invoking the procedure AbLine,
as shown in Step~\ref{AVS:AL_call} of Algorithm~\ref{algo:AVS}.
Again, for each disc absentee in Octant~1 and Octant~2 on $xy$-plane, 
there are two absentee circles---one for the upper hemisphere and another for the lower.
These absentee circles are computed by invoking the procedure AbCircle, 
as shown in Step~\ref{AVS:AC_call} of Algorithm~\ref{algo:AVS}.

\newcommand{\abl}{{\tt AbLine\ }}
\newcommand{\abc}{{\tt AbCircle\ }}

The procedure \abl takes the coordinates $(i_a,k_a)$ of a disc absentee as input.
It also takes a radius $r$ as the third argument, such that $(i_a,k_a)$ lies between $\dc{r}$ and $\dc{r+1}$.
Based on these, it computes the voxels comprising the absentee lines 
$\left\{\al{i'_a,k'_a}:\{|i'_a|\}\cup\{|k'_a|\}=\{i_a,k_a\}\right\}$.

The procedure \abc requires only the coordinates $(i_a,j_a)$ of a disc absentee as input.
If the disc absentee has $i_a=j_a$, then there arises one absentee circle with radius~$i_a$
and center~$(0,j_a,0)$;
otherwise, there are two absentee circles with radius~$i_a$ and $j_a$, and centered at
$(0,j_a,0)$ and $(0,i_a,0)$, respectively.

\begin{algorithm}[!t]
\caption{(AVS) Fixing absentee-voxels in solid sphere}
\label{algo:AVS}
\KwIn{Radius $r$ of a solid sphere}
\KwOut{Set of the absentees}
$\aS{r} \leftarrow \emptyset$\\
{\bf int} $i\leftarrow 0, j\leftarrow r, s\leftarrow 0, w\leftarrow r-1 , h\leftarrow 0,i_a,j_a$\\
{\bf int} $l\leftarrow 2w$\\
\While{$j\geq i$}{
    \Repeat{$s\leq w$}{
	  $s\leftarrow s+2i+1$, $i\leftarrow i+1$
    }
    $i_a\leftarrow i-1, j_a\leftarrow j$ \label{a:runend}\\
    \While{$j_a\geq i_a$}{
	\If{$i_a^2<(2h+1)j_a+h^2$}{
	    $j_a\leftarrow j_a-1$
	}\Else{
	\If{$i_a^2<(2h+1)j_a+(h+1)^2$}{
		$\aS{r}\leftarrow \aS{r} \cup ${\tt AbLine}$(i_a,j_a,j_a+h)$\label{AVS:AL_call}\\
		\If{$i_a=j_a$}{\label{AVS:AC_call}
			$\aS{r}\leftarrow \aS{r} \cup${\tt AbCircle}$(i_a,j_a)$
		}\Else{
			$\aS{r}\leftarrow \aS{r} \cup ${\tt AbCircle}$(i_a,j_a)$\\
			$\aS{r}\leftarrow \aS{r} \cup ${\tt AbCircle}$(j_a,i_a)$
		}
	}
	    $i_a\leftarrow i_a-1$
        }
    }
    $w\leftarrow w+l$, $l\leftarrow l-2$, $j\leftarrow  j-1$, $h\leftarrow h+1$
}
\Return {$\aS{r}$}
\end{algorithm}

\begin{figure}[!t]\center
\includegraphics[width=0.48\textwidth]{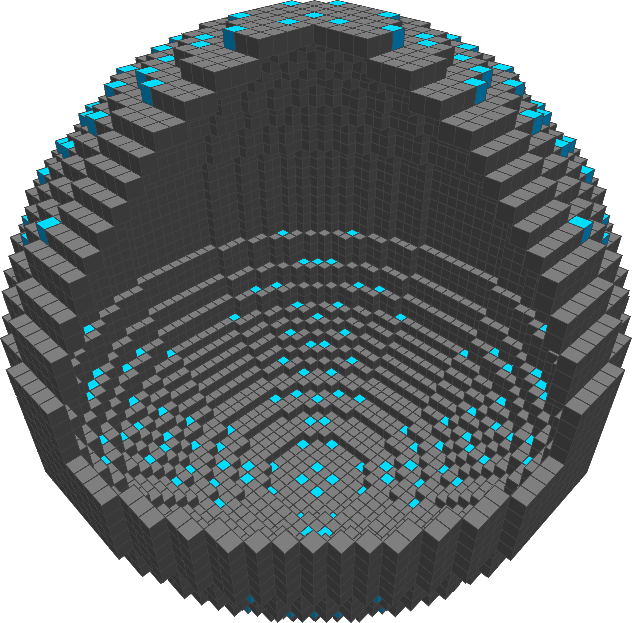}\hfill
\includegraphics[width=0.48\textwidth]{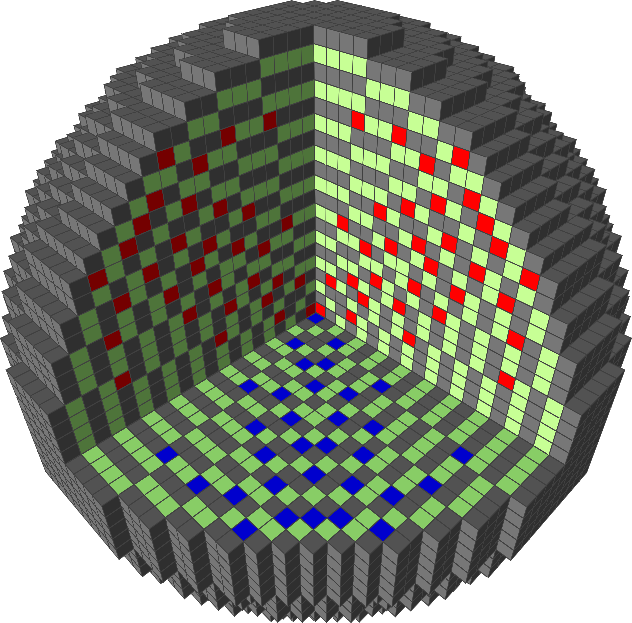}
\caption{Sphere and solid sphere of radius~20 generated by the proposed algorithm.}
\label{fig:sphere20}
\end{figure}

\begin{figure}[!t]\center
\begin{minipage}{.95\linewidth}
\begin{procedure}[H]
\caption{AbLine($i_a,k_a,r$)}
\label{proc:AL}
$A\leftarrow\emptyset$\\
{\bf int} $j_a\leftarrow 0$\\
 \While{$j_a\leq \lfloor\sqrt{r}\rfloor +1$}{
       $A \leftarrow A \cup \{(i'_a,j'_a,k'_a): \{|i'_a|\}\cup\{|k'_a|\}=\{i_a,k_a\} \wedge j_a=|j'_a|\}$\\
	$j_a\leftarrow j_a+1$
 }
\Return{$A$}
\end{procedure}
\end{minipage}\\
\begin{minipage}{.95\linewidth}
\begin{procedure}[H]
\caption{AbCircle($i_a,j_a$)}
\label{proc:AC}
$A\leftarrow\emptyset, r\leftarrow i_a$\\
{\bf int} $i_a\leftarrow 0, k_a\leftarrow r, s\leftarrow 0, w\leftarrow r-1$\\
{\bf int} $l\leftarrow 2w$\\
 \While{$k_a\geq i_a$}{
     \Repeat{$s\leq w$}{
	  $A \leftarrow A \cup \{(i'_a,j_a,k'_a): \{|i'_a|\}\cup\{|k'_a|\}=\{i_a,k_a\} \wedge j_a=|j'_a|\}$\\
 	  $s\leftarrow s+2i_a+1$
 	  $i_a\leftarrow i_a+1$}
     $w\leftarrow w+l$, $l\leftarrow l-2$, $k_a\leftarrow k_a-1$
 }
\Return{$A$}
\end{procedure}
\end{minipage}
\end{figure}

\section{Test Results and Conclusion}
\label{sec:results}

We have implemented Algorithm~\ref{algo:AVH} and Algorithm~\ref{algo:AVS} to generate 
absentee-free spheres and solid spheres.
Figure~\ref{fig:sphere20} shows (absentee-free) instances of a sphere and a solid sphere 
generated by Algorithm~\ref{algo:AVH} for radius~20.

\begin{figure}[h+]\center
\fbox{\includegraphics[scale=.7]{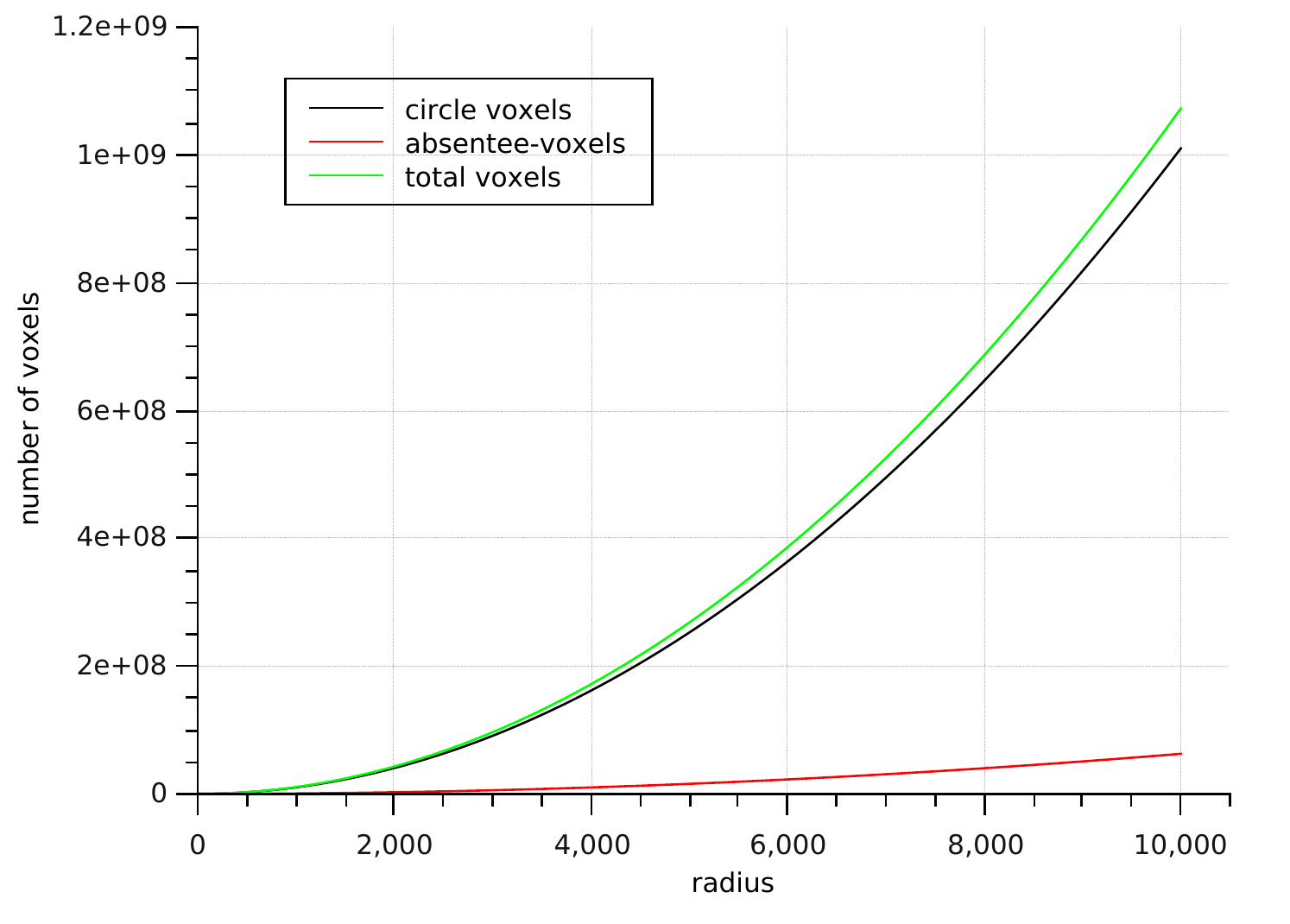}}
\caption{Exact counts of voxels in $\sr{r}$, $\as{r}$, and $\sp{r}$.}
\label{fig:sphere-count}\end{figure}

\begin{figure}[!t]\center
\fbox{\includegraphics[scale=.7]{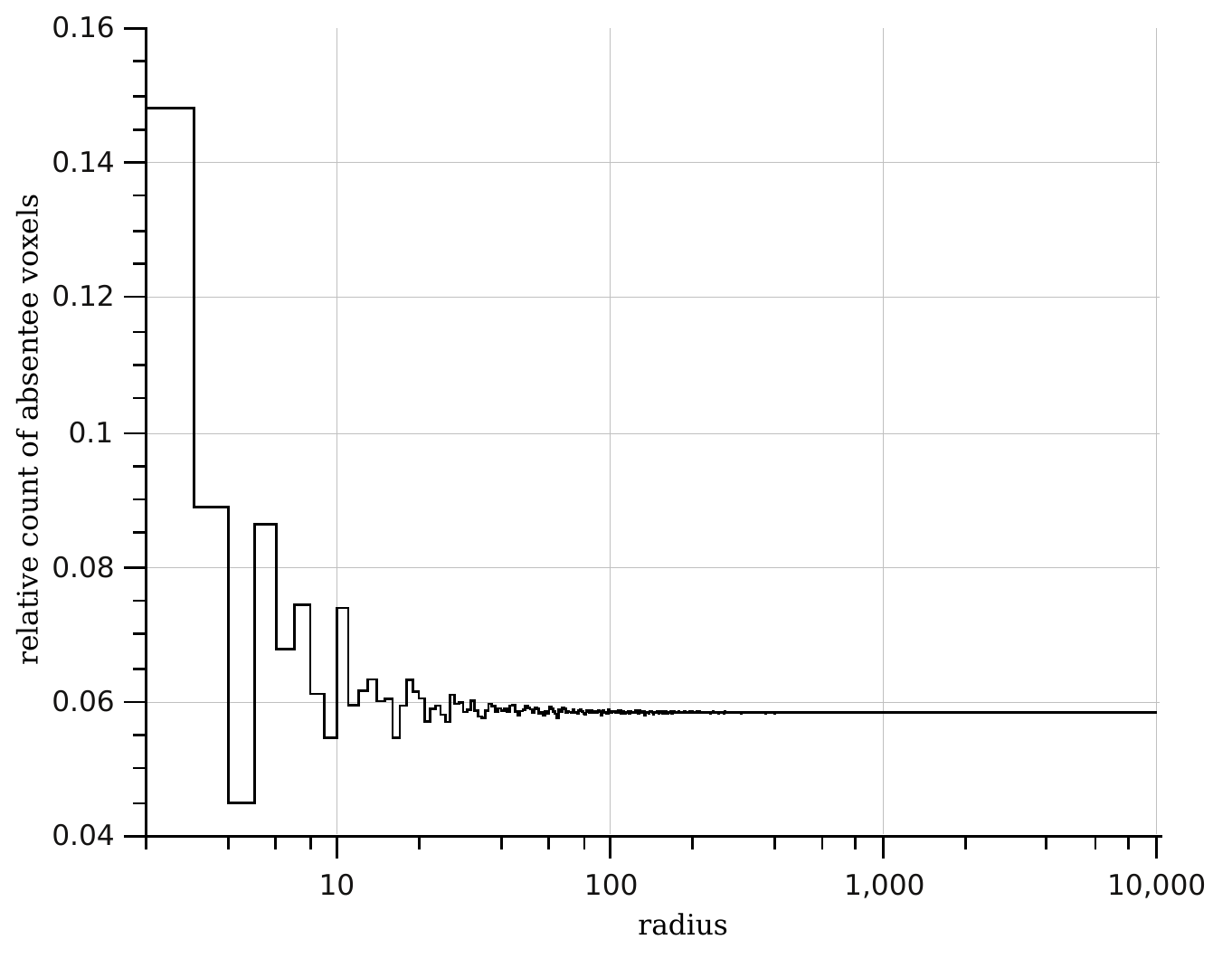}}
\caption{Relative count of absentees versus radius in spheres of revolution.}
\label{fig:sphere-re}
\end{figure}

We have performed experiments to compute the exact counts of absentee-voxels and sphere voxels 
for increasing radius of spheres of revolution.
Table~\ref{tab:sphere} shows the counts of voxels in $\sr{r}$, $\as{r}$, and $\sp{r}$,
for $r$ up~to 10000.
We have also plotted these counts against radius $r$ in Figure~\ref{fig:sphere-count}.
These experimental results reinforce our analytical findings that 
all the three counts have a quadratic dependency on $r$.
The relative counts of absentee-voxels corresponding to digital spheres 
of revolution for radius up~to 10000 are tabulated in Table~\ref{tab:sphere-re} 
and plotted in Figure~\ref{fig:sphere-re}.
In Table~\ref{tab:sphere-re}, the relative count $\as{r}/\sp{r}$ is denoted by $\alpha_r$.
We observe from these data that with the increasing radius, the value of relative count 
for solid sphere tends to $0.058$ approximately.

\begin{table}[!t]\center\footnotesize{
\caption{Exact counts of voxels in $\sr{r}$, $\as{r}$, and $\sp{r}$.}
\scalebox{0.9}{
\fbox{\begin{tabular}{@{\ }c@{\ }c@{\ }}
\begin{tabular}{|@{\ }r@{\ }|@{\ }r@{\ }|@{\ }r@{\ }|@{\ }r@{\ }|@{ }}\hline
$r$& $|\sr{r}|$ &$2|\as{r}|$ & $|\sp{r}|$ \\\hline
       0 &              1 &            0 &             1\\\hline
       1 &              6 &            0 &             6\\\hline
       2 &            46 &            8 &           54\\\hline
       3 &            82 &            8 &           90\\\hline
       4 &          170 &            8 &         178\\\hline
       5 &          254 &          24 &         278\\\hline
       6 &          330 &          24 &         354\\\hline
       7 &          498 &          40 &         538\\\hline
       8 &          614 &          40 &         654\\\hline
       9 &          830 &          48 &         878\\\hline
     10 &        1002 &          80 &       1082\\\hline
     20 &        3978 &        256 &       4234\\\hline
     30 &        8962 &        560 &       9522\\\hline
     40 &      16310 &      1016 &     17326\\\hline
     50 &      25374 &      1592 &     26966\\\hline
     60 &      36438 &      2296 &     38734\\\hline
     70 &      49510 &      3080 &     52590\\\hline
     80 &      64526 &      3992 &     68518\\\hline
     90 &      81582 &      5080 &     86662\\\hline
   100 &    100622 &      6248 &   106870\\\hline
   200 &    404262 &    25104 &    429366\\\hline
   300 &    908250 &    56320 &    964570\\\hline
   400 &  1617026 &  100304 &  1717330\\\hline
   500 &  2524486 &  156608 &  2681094\\\hline
   600 &  3638230 &  225456 &  3863686\\\hline
   700 &  4949282 &  307064 &  5256346\\\hline
   800 &   6461350 & 400768 &  6862118\\\hline
   900 &   8182310 & 507392 &  8689702\\\hline
 1000 & 10097978 & 626304 & 10724282\\\hline
 1100 & 12223938 & 757888 & 12981826\\\hline
\end{tabular}&
\begin{tabular}{|@{\ }r@{\ }|@{\ }r@{\ }|@{\ }r@{\ }|@{\ }r@{\ }|@{ }}\hline
$r$& $|\sr{r}|$ &$2|\as{r}|$ & $|\sp{r}|$ \\\hline
  1200 &     14543190 &     902056 &     15445246\\\hline
  1300 &     17063386 &   1058408 &     18121794\\\hline
  1400 &     19796562 &   1227664 &     21024226\\\hline
  1500 &     22720358 &   1409144 &     24129502\\\hline
  1600 &     25858590 &   1603424 &     27462014\\\hline
  1700 &     29186106 &   1810216 &     30996322\\\hline
  1800 &     32729258 &   2029288 &     34758546\\\hline
  1900 &     36460174 &   2261192 &     38721366\\\hline
  2000 &     40391978 &   2505328 &     42897306\\\hline
  2100 &     44542482 &   2762328 &     47304810\\\hline
  2200 &     48877878 &   3031440 &     51909318\\\hline
  2300 &     53433334 &   3313344 &     56746678\\\hline
  2400 &     58172210 &   3607600 &     61779810\\\hline
  2500 &     63132842 &   3914608 &     67047450\\\hline
  2600 &     68275238 &   4234008 &     72509246\\\hline
  3000 &     90906366 &   5637120 &     96543486\\\hline
  3500 &   123729002 &   7672616 &   131401618\\\hline
  4000 &   161600518 & 10021480 &   171621998\\\hline
  4500 &   204521258 & 12683288 &   217204546\\\hline
  5000 &   252490950 & 15658504 &   268149454\\\hline
  5500 &   305509450 & 18946648 &   324456098\\\hline
  6000 &   363576838 & 22548008 &   386124846\\\hline
  6500 &   426693594 & 26462560 &   453156154\\\hline
  7000 &   494859006 & 30690136 &   525549142\\\hline
  7500 &   568134414 & 35231256 &   603365670\\\hline
  8000 &   646401914 & 40085200 &   686487114\\\hline
  8500 &   729718814 & 45252704 &   774971518\\\hline
  9000 &   818084450 & 50732656 &   868817106\\\hline
  9500 &   911499582 & 56526944 &   968026526\\\hline
10000 & 1009962778 & 62620784 & 1072583562\\\hline
\end{tabular}
\end{tabular}}
}
\label{tab:sphere}
}
\end{table}

\begin{table}[t]\center\footnotesize{
\caption{Relative count of absentees versus radius in spheres of revolution.}
\fbox{\begin{tabular}{@{ }c@{ }c@{ }c@{ }c@{ }}
\begin{tabular}{|@{ }r@{ }|@{ }r@{ }|}\hline
\parbox{8mm}{\center $r$\smallskip}&
\parbox{15mm}{\center $\alpha_r$\smallskip}\\\hline
      2 &         0.148148\\\hline
      3 &         0.088889\\\hline
      4 &         0.044944\\\hline
      5 &         0.086331\\\hline
      6 &         0.067797\\\hline
      7 &         0.074349\\\hline
      8 &         0.061162\\\hline
      9 &         0.054670\\\hline
     10 &         0.073937\\\hline
     11 &         0.059435\\\hline
     12 &         0.061617\\\hline
     13 &         0.063277\\\hline
     14 &         0.060094\\\hline
     15 &         0.060453\\\hline
     16 &         0.054637\\\hline
\end{tabular}&

\begin{tabular}{|@{ }r@{ }|@{ }r@{ }|}\hline
\parbox{8mm}{\center $r$\smallskip}&
\parbox{15mm}{\center $\alpha_r$\smallskip}\\\hline
     17 &         0.059393\\\hline
     18 &         0.063269\\\hline
     19 &         0.061507\\\hline
     20 &         0.060463\\\hline
     30 &         0.058811\\\hline
     40 &         0.058640\\\hline
     50 &         0.059037\\\hline
     60 &         0.059276\\\hline
     70 &         0.058566\\\hline
     80 &         0.058262\\\hline
     90 &         0.058618\\\hline
    100 &         0.058464\\\hline
    120 &         0.058367\\\hline
    140 &         0.058532\\\hline
    160 &         0.058495\\\hline
\end{tabular}&

\begin{tabular}{|@{ }r@{ }|@{ }r@{ }|}\hline
\parbox{8mm}{\center $r$\smallskip}&
\parbox{15mm}{\center $\alpha_r$\smallskip}\\\hline
    180 &         0.058313\\\hline
    200 &         0.058468\\\hline
    300 &         0.058389\\\hline
    400 &         0.058407\\\hline
    500 &         0.058412\\\hline
    600 &         0.058353\\\hline
    700 &         0.058418\\\hline
    800 &         0.058403\\\hline
    900 &         0.058390\\\hline
   1000 &         0.058401\\\hline
   1100 &         0.058381\\\hline
   1200 &         0.058404\\\hline
   1300 &         0.058405\\\hline
   1500 &         0.058399\\\hline
   1600 &         0.058387\\\hline
\end{tabular}&

\begin{tabular}{|@{ }r@{ }|@{ }r@{ }|}\hline
\parbox{8mm}{\center $r$\smallskip}&
\parbox{15mm}{\center $\alpha_r$\smallskip}\\\hline
   1700 &         0.058401\\\hline
   1800 &         0.058382\\\hline
   1900 &         0.058397\\\hline
   2000 &         0.058403\\\hline
   2500 &         0.058386\\\hline
   3000 &         0.058389\\\hline
   3500 &         0.058391\\\hline
   4000 &         0.058393\\\hline
   4500 &         0.058393\\\hline
   5000 &         0.058395\\\hline
   6000 &         0.058396\\\hline
   7000 &         0.058396\\\hline
   8000 &         0.058392\\\hline
   9000 &         0.058393\\\hline
  10000 &         0.058383\\\hline
\end{tabular}
\end{tabular}}
\label{tab:sphere-re}
}
\end{table}

\begin{figure}\center
\fbox{\includegraphics[scale=.7]{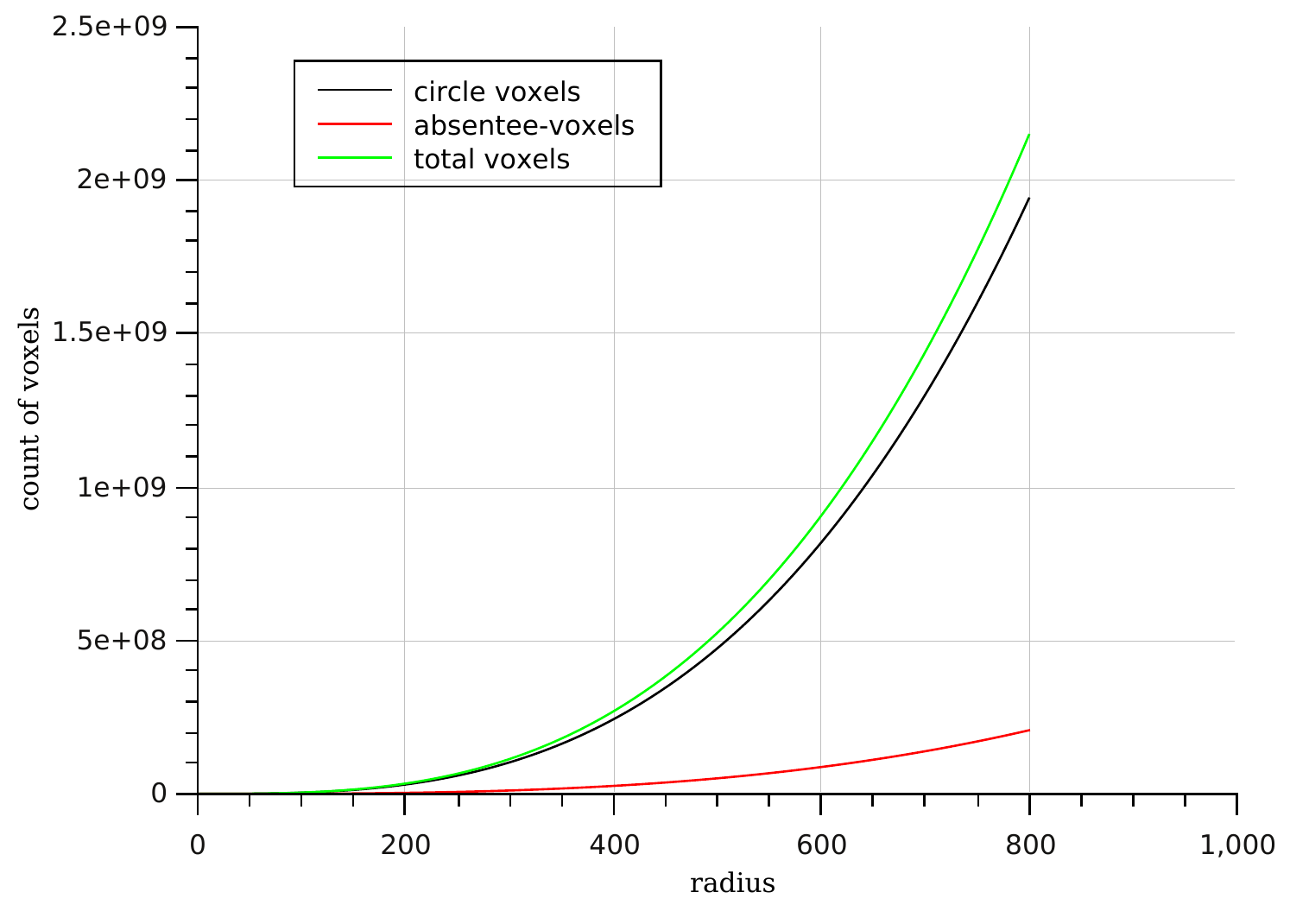}}
\caption{Exact counts of voxels in $\sa{r}$ (circle voxels), $\aS{r}$ (absentee-voxels), and $\ss{r}$
(total voxels).}
\label{fig:sphere-solid-count}\end{figure}

\begin{figure}\center
\fbox{\includegraphics[scale=.7]{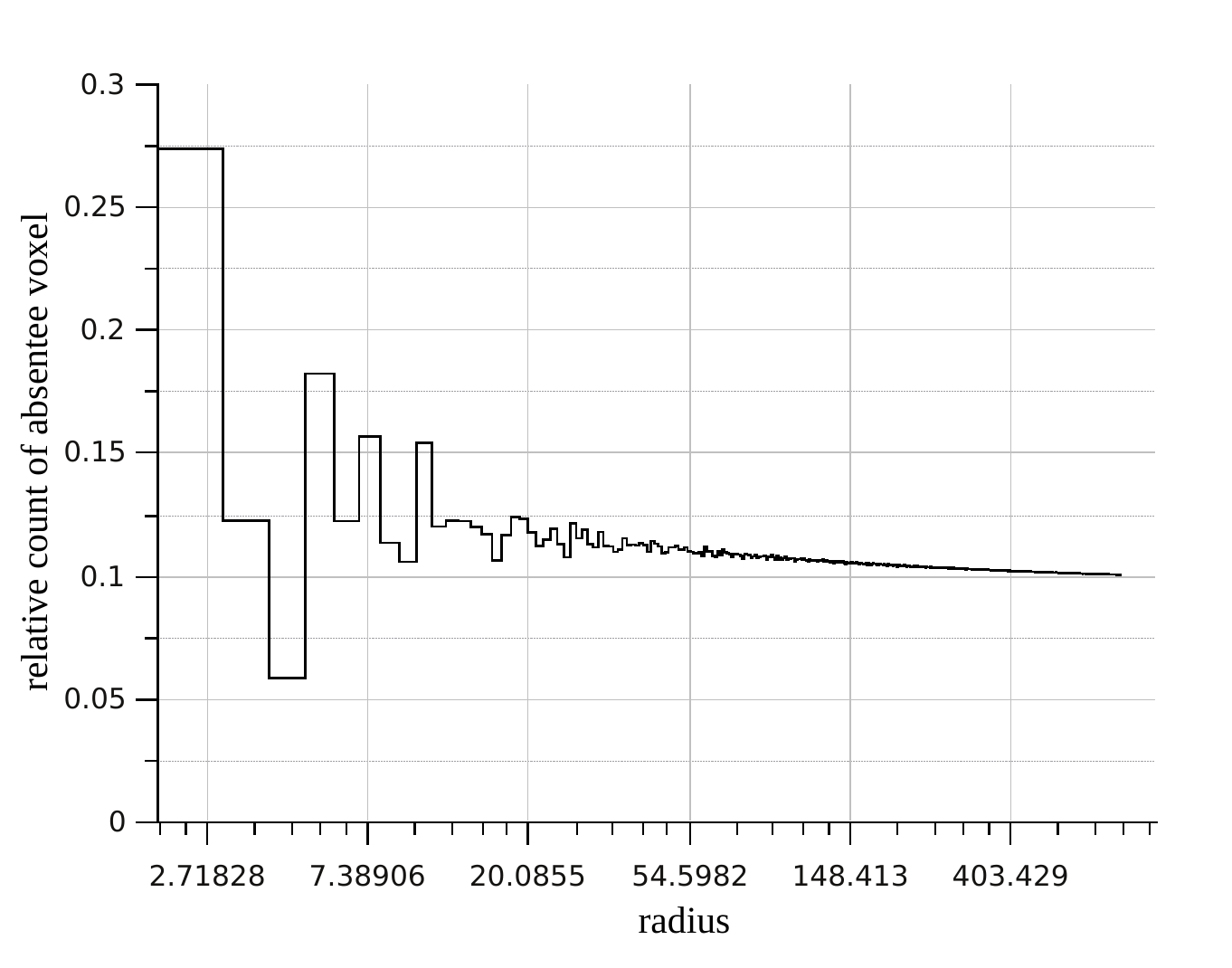}}
\caption{Relative counts of absentee-voxels versus radius in solid spheres of revolution.}
\label{fig:solid-relative-count}
\end{figure}

We have also generated through our experiments the exact counts of absentee-voxels and sphere voxels 
corresponding to solid spheres of revolution.
The counts of voxels in $\sa{r}$, $\aS{r}$, and $\ss{r}$, for $r$ up~to 800,
are shown in Table~\ref{tab:sphere-solid} and plotted in Figure~\ref{fig:sphere-solid-count}.
Similar to the previous set of results, these experimental results also reinforce our analytical 
findings that all the three counts corresponding to the solid sphere have a cubic dependency on $r$.
The relative counts of absentee-voxels corresponding to solid spheres 
of revolution for radius up~to 800 are tabulated in Table~\ref{tab:sphere-soild-re} 
and plotted in Figure~\ref{fig:solid-relative-count}.
We observe from these data that with the increasing radius, the value of relative count tends to 
$0.101$ approximately.

\begin{table}[t]\center\footnotesize{
\caption{Exact counts of voxels in $\sa{r}$, $\aS{r}$, and $\ss{r}$.}
\fbox{\begin{tabular}{@{\ }c@{\ }c@{\ }}
\begin{tabular}{|@{\ }r@{\ }|@{\ }r@{\ }|@{\ }r@{\ }|@{\ }r@{\ }|@{ }}\hline
$r$& $|\sa{r}|$ &$2|\aS{r}|$ & $|\ss{r}|$ \\\hline
	  0 &       	         1 &                    0 &                   1\\\hline
	  1 &     	         7 &                    0 &                   7\\\hline
	  2 &      	        53 &                   20 &                  73\\\hline
	  3 &                  143 &                   20 &                 163\\\hline
          4 &                  321 &                   20 &                 341\\\hline
          5 &                  591 &                  132 &                 723\\\hline
          6 &                  945 &                  132 &                1077\\\hline
          7 &                 1483 &                  276 &                1759\\\hline
          8 &                 2153 &                  276 &                2429\\\hline
          9 &                 3039 &                  360 &                3399\\\hline
         10 &                 4121 &                  752 &                4873\\\hline
         20 &                31377 &                 4192 &               35569\\\hline
         30 &               104321 &                13144 &              117465\\\hline
         40 &               245349 &                31412 &              276761\\\hline
         50 &               477061 &                60436 &              537497\\\hline
         60 &               821805 &               103604 &              925409\\\hline
         70 &              1303165 &               159636 &             1462801\\\hline
\end{tabular}&

\begin{tabular}{|@{\ }r@{\ }|@{\ }r@{\ }|@{\ }r@{\ }|@{\ }r@{\ }|@{ }}\hline
$r$& $|\sa{r}|$ &$2|\aS{r}|$ & $|\ss{r}|$ \\\hline
         80 &              1941629 &               233828 &             2175457\\\hline
         90 &              2761237 &               333428 &             3094665\\\hline
        100 &              3785733 &               452052 &             4237785\\\hline
        150 &             12749489 &              1508868 &            14258357\\\hline
        200 &             30196125 &              3528744 &            33724869\\\hline
        250 &             58952525 &              6810356 &            65762881\\\hline
        300 &            101848409 &             11688640 &           113537049\\\hline
        350 &            161726089 &             18514264 &           180240353\\\hline
        400 &            241406453 &             27530128 &           268936581\\\hline
        450 &            343714485 &             39030584 &           382745069\\\hline
        500 &            471497269 &             53389448 &           524886717\\\hline  
        550 &            627583253 &             70890036 &           698473289\\\hline   
        600 &            814799465 &             91803032 &           906602497\\\hline
        650 &           1035980249 &            116498872 &          1152479121\\\hline
        700 &           1293980265 &            145396532 &          1439376797\\\hline  
        750 &           1591598569 &            178467668 &          1770066237\\\hline
        800 &           1931678709 &            216171360 &         2147850069\\\hline  
\end{tabular}
\end{tabular}}
\label{tab:sphere-solid}
}
\end{table}

\begin{table}[t]\center\footnotesize{
\caption{Relative counts of absentee-voxels versus radius in solid spheres of revolution.}
\fbox{\begin{tabular}{@{ }c@{ }c@{ }c@{ }c@{ }}
\begin{tabular}{|@{ }r@{ }|@{ }r@{ }|}\hline
\parbox{8mm}{\center $r$\smallskip}&
\parbox{15mm}{\center $\alpha_r$\smallskip}\\\hline
          2 &                0.27397\\\hline
          3 &                0.12270\\\hline
          4 &                0.05865\\\hline
          5 &                0.18257\\\hline
          6 &                0.12256\\\hline
          7 &                0.15691\\\hline
          8 &                0.11363\\\hline
          9 &                0.10591\\\hline
         10 &                0.15432\\\hline
         11 &                0.12030\\\hline
         12 &                0.12264\\\hline
         13 &                0.12249\\\hline
         14 &                0.12018\\\hline
         15 &                0.11719\\\hline
\end{tabular}&

\begin{tabular}{|@{ }r@{ }|@{ }r@{ }|}\hline
\parbox{8mm}{\center $r$\smallskip}&
\parbox{15mm}{\center $\alpha_r$\smallskip}\\\hline
         16 &                0.10654\\\hline
         17 &                0.11685\\\hline
         18 &                0.12414\\\hline
         19 &                0.12345\\\hline
         20 &                0.11786\\\hline
         30 &                0.11190\\\hline
         40 &                0.11350\\\hline
         50 &                0.11244\\\hline
         60 &                0.11195\\\hline
         70 &                0.10913\\\hline
         80 &                0.10748\\\hline
         90 &                0.10774\\\hline
        100 &                0.10667\\\hline
        110 &                0.10661\\\hline
\end{tabular}&

\begin{tabular}{|@{ }r@{ }|@{ }r@{ }|}\hline
\parbox{8mm}{\center $r$\smallskip}&
\parbox{15mm}{\center $\alpha_r$\smallskip}\\\hline
        120 &                0.10654\\\hline
        130 &                0.10620\\\hline
        140 &                0.10606\\\hline
        150 &                0.10582\\\hline
        160 &                0.10523\\\hline
        170 &                0.10482\\\hline
        180 &                0.10482\\\hline
        190 &                0.10463\\\hline
        200 &                0.10463\\\hline
        220 &                0.10403\\\hline
        240 &                0.10368\\\hline
        260 &                0.10361\\\hline
        280 &                0.10328\\\hline
        300 &                0.10295\\\hline
\end{tabular}&

\begin{tabular}{|@{ }r@{ }|@{ }r@{ }|}\hline
\parbox{8mm}{\center $r$\smallskip}&
\parbox{15mm}{\center $\alpha_r$\smallskip}\\\hline
        320 &                0.10300\\\hline
        340 &                0.10283\\\hline
        360 &                0.10250\\\hline
        380 &                0.10233\\\hline
        400 &                0.10237\\\hline
        420 &                0.10219\\\hline
        450 &                0.10198\\\hline
        500 &                0.10172\\\hline
        550 &                0.10149\\\hline
        600 &                0.10126\\\hline
        650 &                0.10109\\\hline
        700 &                0.10101\\\hline
        750 &                0.10083\\\hline
        800 &                0.10065\\\hline
\end{tabular}
\end{tabular}}
\label{tab:sphere-soild-re}
}
\end{table}

\enlargethispage*{3mm}

The above test results and their theoretical analysis indicate that the ratio of the absentee-voxels 
to the total number of voxels tends to a constant for large radius.
The knowledge of geometric distributions of absentee-voxels is shown to be useful for an algorithmic
construction of a digital sphere. Although an asymptotic tight bound for the count of absentees has been
provided here, the determination of a closed-form solution of the exact count of absentees for a given radius,
still remains an open problem. The characterization of these absentees requires further in-depth analysis,
especially if we want to generate a solid digital sphere with a set of concentric digital spheres. Apart from
spheres, the generation of various other types of surfaces that are devoid of any absentee-voxels, will also
have many applications in 3D imaging and graphics, such as the creation of interesting pottery designs, as
reported recently~\cite{kumar_11}.


\bibliographystyle{abbrv}
\bibliography{sahadev,ranita}

\end{document}